\documentclass[twocolumn,aps,preprintnumbers,pra,showpacs%
,amsfonts,amsmath,amssymb,superscriptaddress,floatfix,dblfloatfix%
,10pt,lettersize]{revtex4}

\usepackage[dvips]{graphicx}
\usepackage[tight]{subfigure}
\usepackage{bm}

\graphicspath{{eps/}}

\allowdisplaybreaks[1]

\newcommand{\hatd}[1]{\hat{#1}^\dagger}
\newcommand{\avg}[1]{\langle{#1}\rangle}

\newcommand{\ket}[1]{|{#1}\rangle}

\newcommand{\realnum}{\mathbb{R}}
\newcommand{\compnum}{\mathbb{C}}

\newcommand{\affA}{%
	Department of Applied Physics, School of Engineering, 
        The University of Tokyo,\\
	7-3-1 Hongo, Bunkyo-ku, Tokyo 113-8656, Japan}

\newcommand{\affB}{%
	Department of Optics, Palack\'{y} University, 17. listopadu 1192/12, 772 07 Olomouc,
    Czech Republic}

\begin{document} 

\title{Demonstration of reversible phase-insensitive optical amplifier}

\date{\today}

\author{Jun-ichi Yoshikawa}
\affiliation{\affA}
\author{Yoshichika Miwa}
\affiliation{\affA}
\author{Radim Filip}
\affiliation{\affB}
\author{Akira Furusawa}
\affiliation{\affA}

\begin{abstract}
We experimentally demonstrate phase-insensitive linear optical amplification which preserves the idler at the output. 
Since our amplification operation is unitary up to small excess noise, it is reversible beyond the classical limit. 
The entanglement between the two output modes is the resource for the reversibility. 
The amplification gain of $G=2.0$ is demonstrated. 
In addition, combining this amplifier with a beamsplitter, we also demonstrate approximate cloning of coherent states where an anticlone is present. 
We investigate the reversibility by reconstructing the initial state from the output correlations, and the results are slightly beyond the cloning limit. 
Furthermore, full characterization of the amplifier and cloner is given by using coherent states with several different mean values as inputs. 
Our amplifier is based on linear optics, offline preparation of nonclassical ancillas, and homodyne measurements followed by feedforward. 
Squeezed states are used as the ancillas, and nonlinear optical effects are exploited only for their generation. 
The ancillas introduce nonclassicality into the amplifying operation, making entanglement at the output. 
\end{abstract}


\maketitle

\section{Introduction}

Quantum optics is governed by rules imposed by commutation relations which have to be kept during time evolution.
Optical amplification is no exception to this story. 
Typically, the amplified output is suffered from inevitable excess noise. 
This limitation is quantum-mechanically imposed, thus does not depend on the specific realization methods. 
Caves classified general linear amplification into phase-insensitive amplification (PIA) and phase-sensitive amplification (PSA)~\cite{Caves(1982):PRD}. 
He also systematically derived the quantum limit of excess noise for such general linear amplification with arbitrary gain from the requirement to preserve commutation relations. 
This excess noise originates from quantum fluctuations in the auxiliary system required to keep energy conservation. 

We concentrate on PIA, supposing the target of amplification to be optical wave amplitude of a single mode, which is denoted by the term ``signal''. 
Classical counterpart of PIA is a conversion of arbitrary complex wave amplitude $\alpha\in\compnum$ into $\sqrt{G}\alpha$, where $G\ge1$ is the gain of amplification.
As is found in ordinary textbooks, annihilation operators in quantum optics correspond to complex amplitudes in classical optics. 
Therefore, we describe the amplifying process by the transformation of annihilation operators. 
Quantum-mechanically optimal PIA in the sense that the excess noise is minimized can be achieved by the following transformation~\cite{Caves(1982):PRD}: 
\begin{align}
\hat{a}_\text{sig}^\text{out}=
\sqrt{G}\,\hat{a}_\text{sig}^\text{in}+e^{i\theta}\sqrt{G-1}\,(\hat{a}_\text{idl}^\text{in})^\dagger,
\label{eq:PiaSingle}
\end{align}
where $\hat{a}_\text{sig}^\text{in}$ and $\hat{a}_\text{sig}^\text{out}$ are the signal mode's annihilation operators before and after the amplification, respectively. 
There is an extra term $e^{i\theta}\sqrt{G-1}\,(\hat{a}_\text{idl}^\text{in})^\dagger$ which is introduced in order to meet the commutation relation of $[\hat{a}_\text{sig},\hat{a}_\text{sig}^\dagger]=1$ for both the input and output signal modes. 
Here, $\theta\in\realnum$ is an arbitrary phase factor, and $\hat{a}_\text{idl}^\text{in}$ is another mode's annihilation operator in the auxiliary system. 
Throughout this paper, the ancilla mode represented by $\hat{a}_\text{idl}^\text{in}$ is denoted by the term ``idler'' and distinguished from other ancilla modes. 
Eq.~\eqref{eq:PiaSingle} becomes the input-output relation of optimal PIA when the idler is in a vacuum state. 
The quantum fluctuation of the idler contaminates the amplified signal. 
This is the inevitable excess noise of PIA. 
Note that this penalty prevents amplification from being a loophole of the uncertainty relation in joint measurements~\cite{Authurs(1965):BSTJ,Authurs(1988):PRL}. 
At the limit of high-gain amplification, we can see the famous $3$~dB cost of the noise figure for PIA of coherent states. 
In addition to this intrinsic excess noise, further nonintrinsic excess noise may be caused by other ancilla modes in nonoptimal PIA. 

There are numerous practical realizations of optical amplification. 
Doped fiber amplifiers (DFAs) and semiconductor optical amplifiers (SOAs) utilize stimulated emissions~\cite{Shimoda(1957):JPSJ}, and Raman amplifiers (RAs) and optical parametric amplifiers (OPAs) utilize nonlinear optical effects. 
In principle, there is no quantum-mechanical reason to prevent these realizations from achieving the optimal PIA in the form of Eq.~\eqref{eq:PiaSingle}. 
However, the real devices with current technology are accompanied by further excess noises. 

Recently, PIA operating almost at the optimal level is experimentally demonstrated by Josse \textit{et al.} by utilizing feedforward~\cite{Josse(2006):PRL}. 
The reason for the high efficiency of Josse's PIA is that it does not require inefficient nonclassical operations or nonclassical ancillas. 
It uses a vacuum state as an ancilla which is present everywhere, and linear optics and homodyne measurements followed by feedforward which are highly efficient. 

Although Josse's PIA is a good attainment, it is not the end of the story. 
The signal transformation in Eq.~\eqref{eq:PiaSingle} is an irreversible thermalizing process. 
Complete PIA should have \textit{unitary} realization on an expanded Hilbert space. 
In order to unitarize PIA, two-mode description is sufficient. 
The full input-output relation becomes as follows: 
\begin{subequations}\label{eq:PiaUnitary}
\begin{align}
\hat{a}_\text{sig}^\text{out}= &
\sqrt{G}\,\hat{a}_\text{sig}^\text{in}+e^{i\theta}\sqrt{G-1}\,(\hat{a}_\text{idl}^\text{in})^\dagger, 
\label{seq:PiaSignal}\\
\hat{a}_\text{idl}^\text{out}= &
\sqrt{G}\,\hat{a}_\text{idl}^\text{in}+e^{i\theta}\sqrt{G-1}\,(\hat{a}_\text{sig}^\text{in})^\dagger.
\label{seq:PiaIdler}
\end{align}
\end{subequations}
Note that the roles of the signal and idler are symmetric in this relation. 

The significance of unitarization must be the reversibility. 
The inverse transformation is easily derived when we take notice of the fact that Eq.~\eqref{eq:PiaUnitary} is equivalent to two-mode squeezing operation. 
A two-mode squeezing operation parametrized by $(G,\theta)$ is canceled by another two-mode squeezing operation where the squeezing direction is opposite, i.e., $(G,\theta+\pi)$. 
Nonetheless, in many amplification schemes including Josse's experimental demonstration~\cite{Josse(2006):PRL}, the idler output is lost in the inextractable environment, making the process irreversible. 

In order to realize idler-preserving and close-to-optimal PIA, we require some nonclassicality for the amplifier. 
This is contrastive to Josse's idler-nonpreserving PIA which does not require any nonclassicality. 
A typical strategy to introduce nonclassicality into feedforward-based quantum circuits is to use nonclassical states as ancillas. 
Continuous-variable (CV) quantum teleportation~\cite{Furusawa(1998):Science} and CV error correction~\cite{Aoki(2009):NatPhys} are good examples. 
In these examples, squeezed states are used as ancillas that support the performance beyond the classical limit, and the complex operations after the state preparation stage are efficiently implemented by linear optics. 

In this paper, by employing the feedforward-based scheme proposed in Ref.~\cite{Filip(2005):PRA}, we demonstrate PIA of coherent states which preserve the idler output.  
The scheme basically relies on linear optics including homodyne measurements and feedforward. 
Squeezed vacuum states are used as ancillas, which inject nonclassicality into our PIA. 
Only for generating the nonclassical ancilla states, we resort to nonlinear optical effects.
Our demonstration is for the amplification gain of $G=2.0$, which is tuned via passive optical devices and feedforward electric circuits. 
Combining PIA for $G=2.0$ with a half beamsplitter, we also demonstrate $1\to2$ approximate cloning of coherent states, where an ``anticlone'' remains at the output. 
(Anticlone will be explained in Sec.~\ref{sec:Clone}.) 
In principle, our amplifier and cloner becomes quantum-mechanically optimum at the limit of infinite squeezing of the ancillas. 
For the case of finite squeezing, as is the real situation in experiments, further excess noise invades in accordance to the level of the squeezing. 
However, the degradation is small enough to retain nonclassical features. 
The behaviors of our amplifier and cloner are fully characterized by using several coherent states as inputs. 
Furthermore, we also pay much attention to the output correlations, because nonclassical properties clearly appear in them. 
For the PIA experiment, we check the Einstein-Podolsky-Rosen (EPR) correlation between the signal and idler outputs. 
For the cloning experiment, we check bipartite entanglement between each clone and the anticlone, which as a whole proves tripartite entanglement of class I~\cite{Giedke(2001):PRA}.
Moreover, for both experiments, the reversibility is investigated from the output correlations. 

Our idler-preserving PIA is significant in several respects. 
First of all, the reversibility will pave the way to new schemes. 
Recently, there is a proposal of a CV quantum interface that enables in principle a unit fidelity of transfer using such reversible PIA~\cite{Radim(2009):PRA}. 
Moreover, the reversibility in cloning is also advantageous. 
Cloning of unknown states is distribution of information, and its reversibility reserves the option to recover the distributed fragments of the information. 
This will be further discussed in Sec.~\ref{sec:Clone}. 
Secondly, our PIA would have some applications as two-mode squeezing operation.
Note that one-mode squeezing operation is already demonstrated successfully in Ref.~\cite{Yoshikawa(2007):PRA} with similar approach. 

In this introduction, PIA has been described together with a brief historical review.
Especially, the nonclassical property of PIA is discussed, which is obscure in many amplification processes because the idler output is lost in the inextractable environment. 
The subsequent contents of this paper are as follows. 
In Sec.~\ref{sec:FfPia}, feedforward-based PIA is described, explicitly showing the excess noise due to finite squeezing of ancillas. 
In Sec.~\ref{sec:Clone}, CV quantum state cloning and its connection with PIA are described. 
In Sec.~\ref{sec:SetUp}, the experimental setup is described. 
In Sec.~\ref{sec:ResultsPia}, the experimental results for PIA of coherent states with $G=2.0$ are shown.
In Sec.~\ref{sec:ResultsClone}, the experimental results for $1\to2$ approximate cloning of coherent states are shown. 
In Sec.~\ref{sec:Summary}, our experimental achievements are summarized.

\section{Feedforward-based Amplifier}
\label{sec:FfPia}

In our definition, feedforward means that the operations after some measurements are changed depending on the measurement outcomes which in general are obtained randomly. 
In particular, in this paper it indicates phase space displacement operations whose amounts are proportional to the results of homodyne measurements. 

We know two specific schemes for feedforward-based PIA that preserves the idler at the output. 
One scheme is proposed by Filip \textit{et al.}\ in Ref.~\cite{Filip(2005):PRA}, in which PIA is composed of two feedforward-based single-mode squeezers proposed in the same paper. 
The other scheme is proposed by Josse \textit{et al.}\ in Ref.~\cite{Josse(2006):PRL} as a modification of the idler-nonpreserving PIA. 
Note that Josse's idler-preserving PIA is just a theoretical proposal and the idler-nonpreserving PIA alone is experimentally demonstrated. 

Both of Filip's scheme and Josse's scheme rely on linear optics including homodyne measurements and feedforward, and require offline-prepared nonclassical states as ancillas. 
Moreover, in both schemes, the gain of amplification is accurately and stably determined via the choice of passive optical devices and correspondingly feedforward gains. 
As for the nonclassical ancillas, Filip's scheme requires two single-mode squeezed states, on the other hand, Josse's scheme requires a two-mode squeezed state. 
Since two single-mode squeezed states can be converted to a two-mode squeezed state and vice versa by a half beamsplitter interaction, the amounts of the nonclassical resources required for the two distinct schemes are the same. 

For both schemes, the feedforward-based PIA coincides with the quantum-mechanically optimal PIA only at the limit of infinite squeezing of the ancillas. 
For the case of finite squeezing, excess noise contaminates the output to some extent. 
Note that this is a common matter of feedforward-based CV deterministic processing~\cite{Furusawa(1998):Science,Aoki(2009):NatPhys,Ukai(2010):QPh}. 
The difference between the two schemes proposed by Filip and Josse solely arises in this excess noise. 
For Filip's scheme, it appears symmetrically in the signal and idler outputs. 
On the other hand, for Josse's scheme, it appears only in the idler output. 
The better choice between different schemes depends on the specific application. 

We have chosen the symmetrical one. 
In the demonstration in Sec.~\ref{sec:ResultsPia}, we confirm the symmetry of PIA by swapping the roles of the signal and idler.

Fig.~\ref{sfig:AmpSchematic} shows the schematic of our PIA, from which the symmetry of the signal and idler is obvious. 
Its details will be described in Sec.~\ref{sec:SetUp}. 
Here we give the input-output relation. 
In the following, the quadrature phase amplitudes of each optical mode are denoted by $\hat{x}$ and $\hat{p}$, which correspond to the real and imaginary parts of the mode's annihilation operator $\hat{a}$, i.e., $\hat{a}=\hat{x}+i\hat{p}$. 
The phase factor $\theta$ in Eq.~\eqref{eq:PiaUnitary} can be arbitrarily changed by pre- and post-processing of phase rotation of the idler. 
Therefore, we consider the case of $\theta=0$ without loss of generality. 
Explicitly showing the excess noise coming from finitely squeezed ancillas, the input-output relation becomes as follows~\cite{Filip(2005):PRA}: 
\begin{subequations}
\begin{align}
\!\!\hat{x}_1^\text{out}\! = &
\tfrac{1}{2}\bigl(\tfrac{1}{\sqrt{\!R}}\!+\!\sqrt{\!R}\bigr)\hat{x}_1^\text{in}
\!+\!\tfrac{1}{2}\bigl(\tfrac{1}{\sqrt{\!R}}\!-\!\sqrt{\!R}\bigr)\hat{x}_2^\text{in}
\!-\!\sqrt{\tfrac{1-R}{2}}\hat{x}_\text{A}^\text{out}\!,\! \\
\!\!\hat{p}_1^\text{out}\! = &
\tfrac{1}{2}\bigl(\tfrac{1}{\sqrt{\!R}}\!+\!\sqrt{\!R}\bigr)\hat{p}_1^\text{in}
\!-\!\tfrac{1}{2}\bigl(\tfrac{1}{\sqrt{\!R}}\!-\!\sqrt{\!R}\bigr)\hat{p}_2^\text{in}
\!+\!\sqrt{\tfrac{1-R}{2}}\hat{p}_\text{B}^\text{out}\!,\! \\
\!\!\hat{x}_2^\text{out}\! = &
\tfrac{1}{2}\bigl(\tfrac{1}{\sqrt{\!R}}\!+\!\sqrt{\!R}\bigr)\hat{x}_2^\text{in}
\!+\!\tfrac{1}{2}\bigl(\tfrac{1}{\sqrt{\!R}}\!-\!\sqrt{\!R}\bigr)\hat{x}_1^\text{in}
\!+\!\sqrt{\tfrac{1-R}{2}}\hat{x}_\text{A}^\text{out}\!,\! \\
\!\!\hat{p}_2^\text{out}\! = &
\tfrac{1}{2}\bigl(\tfrac{1}{\sqrt{\!R}}\!+\!\sqrt{\!R}\bigr)\hat{p}_2^\text{in}
\!-\!\tfrac{1}{2}\bigl(\tfrac{1}{\sqrt{\!R}}\!-\!\sqrt{\!R}\bigr)\hat{p}_1^\text{in}
\!+\!\sqrt{\tfrac{1-R}{2}}\hat{p}_\text{B}^\text{out}\!.\!
\end{align}
\end{subequations}
The subscripts `1' and `2' represent the two main modes.
They correspond to the signal and idler, though we do not specify which is which because the relation is symmetric. 
$\hat{x}_\text{A}$ and $\hat{p}_\text{B}$ denote the squeezed quadratures of the two ancilla modes. 
At the limit of infinite squeezing, these terms vanish, and the transformation above strictly coincides with the optimal PIA. 
The amplification gain $G$ is determined via one parameter $R$, which is a common reflectivity of two beamsplitters \mbox{BS-A} and \mbox{BS-B} in Fig.~\ref{sfig:AmpSchematic}, with the relation of
\begin{align} 
G=\tfrac{1}{4}\bigl(\tfrac{1}{\sqrt{R}}+\sqrt{R}\bigr)^2.
\label{eq:Gain}
\end{align}
One-to-one correspondence of $1\le{G}<\infty$ and $0<R\le1$ is easily checked. 
Note that the feedforward gain is also parameterized by $R$. 
It is chosen so that the antisqueezed noises from the ancillas are canceled out at the output. 

For the demonstration of $G=2$, The value of $R$ should be chosen as $3-2\sqrt{2}\approx0.17$.
The resulting input-output relation becomes as follows: 
\begin{subequations}\label{eq:in-out_G2}
\begin{align}
\hat{x}_1^\text{out} = &
\sqrt{2}\,\hat{x}_1^\text{in}
+\hat{x}_2^\text{in}
-\sqrt{\sqrt{2}-1}\,\hat{x}_\text{A}^\text{out}, 
\label{}\\
\hat{p}_1^\text{out} = &
\sqrt{2}\,\hat{p}_1^\text{in}
-\hat{p}_2^\text{in}
+\sqrt{\sqrt{2}-1}\,\hat{p}_\text{B}^\text{out}, \\
\hat{x}_2^\text{out} = &
\sqrt{2}\,\hat{x}_2^\text{in}
+\hat{x}_1^\text{in}
+\sqrt{\sqrt{2}-1}\,\hat{x}_\text{A}^\text{out}, \\
\hat{p}_2^\text{out} = &
\sqrt{2}\,\hat{p}_2^\text{in}
-\hat{p}_1^\text{in}
+\sqrt{\sqrt{2}-1}\,\hat{p}_\text{B}^\text{out}.
\end{align}
\end{subequations}

\section{Quantum State Cloning}
\label{sec:Clone}

It is known as the no-cloning theorem that an unknown quantum state $\ket{\psi}$ cannot be perfectly duplicated as $\ket{\psi}\ket{\psi}$~\cite{Wootters(1982):Nature}. 
However, approximate cloning is possible, which can go beyond some classical limit in general. 

In this section, we will discuss CV cloning, and make its connection with PIA. 
Furthermore, its reversibility is discussed by introducing the notion of anticlone. 
In general, cloning can be described as a unitary operation supported by ancilla systems. 
The ancilla output system generally depends on the cloned state, from which anticlones are obtained. 
We show the equations for $1\to2$ cloning, which corresponds to the experimental demonstration in Sec.~\ref{sec:ResultsClone}. 
However, PIA allows general ${K}\to{L}$ cloning in principle, which will be described in Appendix~\ref{sec:GenCln}. 
Here, the notation ${K}\to{L}$ means that $L$ clones are created from $K$ identical originals. 

First, we would like to say that pragmatic cloning for CV is not the universal cloning~\cite{Braunstein(2001):PRA} with respect to the infinite-dimensional Hilbert space, because it is an unnatural situation that all the states in the noncompact space appears with equal probability. 
In general, the choice of the appropriate cloner depends on how the information is embedded in the infinite-dimensional Hilbert space. 

The typical situation is that the CV information is embedded as a displacement on some quantum state $\ket{\psi}$. 
Here, $\ket{\psi}$, which we refer to as a core state, is either known or unknown. 
Then, the set of possible original states is $S=\{\hat{D}(x_\text{d},p_\text{d})\ket{\psi}\!\mid\!(x_\text{d},p_\text{d})\in\realnum^2\}$, where $\hat{D}(x_\text{d},p_\text{d})\equiv\exp[-2i(x_\text{d}\hat{p}-p_\text{d}\hat{x})]$ is the displacement operator. 
[The probability density $p(x_\text{d},p_\text{d})$ is omitted because we consider for simplicity the case where $(x_\text{d},p_\text{d})$ is uniformly distributed.] 
As a special case of this, the set $S$ becomes all coherent states when the core state $\ket{\psi}$ is known to be a vacuum state. 
This way of embedding is found in ordinary CV quantum key distribution (QKD) protocols~\cite{Grosshans(2002):PRL}. 

For such protocols, the role of cloning is distribution of the information, rather than duplication of a quantum state. 
Therefore, the measure of the cloning precision should be related to the estimation of $(x_\text{d},p_\text{d})$, instead of the traditional fidelity. 
Furthermore, asymmetric cloner is significant as well as symmetric cloner. 
Arbitrary share ratio of the information is achieved by cloner with tunable asymmetry. 

We suppose a simple picture of cloning where some noise is added to the original state as the penalty of cloning. 
Then, the quality of cloning is totally determined by this noise. 
For simplicity, we impose rotational symmetry on the noise added to each clone.  
This is naturally justified when the core state $\ket{\psi}$ is either known to be symmetric or unknown. 
The added noise is characterized by its variance $n_k\equiv({\Delta}x_{\text{cln-}k}^\text{noise})^2+({\Delta}p_{\text{cln-}k}^\text{noise})^2$~\cite{Fiurasek(2007):PRA}, where $k\in\{1,2\}$ for $1\to2$ cloning. 
Note that $n_k$ corresponds to the mean photon number of thermalization in the \mbox{$k$-th} clone. 

The variances $n_k$ are directly connected to the mean square errors in the estimation of $(x_\text{d},p_\text{d})$. 
Therefore, a measure can be constructed from them. 
Given the desired asymmetry, the cost function is determined~\cite{Fiurasek(2007):PRA}: 
\begin{gather}
C(n_1,n_2)=c_1n_1+c_2n_2. 
\label{eq:ClnCostFunc}
\end{gather}
The positive parameters $c_k$ determine the asymmetry. 
($c_1=c_2$ corresponds to symmetric cloning.) 
The cloner that minimize the cost function is the optimal. 

It is obvious that the optimal cloner is Gaussian when the cost function is set as a function of the noise variances as in Eq.~\eqref{eq:ClnCostFunc}. 
In order to minimize it, the ancillas that support cloning are chosen in minimum uncertainty states, which are Gaussian. 
This contrasts with the evaluation by the fidelity. 
Non-Gaussian cloning can slightly go beyond the Gaussian fidelity for coherent states~\cite{Cerf(2005):PRL}. 
We emphasize that the evaluation by the variances is more practical. 
It is pointed out that optimal attack in QKD is Gaussian~\cite{Grosshans(2004):PRL,Leverrier(2010):PRA}. 

There is a restriction on the excess noises $n_k$ which is imposed by quantum mechanics~\cite{Cerf(2000):PRL,Fiurasek(2001):PRL}: 
\begin{align}
n_1n_2\ge(1/2)^2.
\label{eq:ClnNoiseIneq}
\end{align} 
The optimal cloner with respect to the cost function in Eq.~\eqref{eq:ClnCostFunc} satisfies the equality in Eq.~\eqref{eq:ClnNoiseIneq} by necessity. 
This noise penalty comes from consistency with the uncertainty relation. 
The attainable information of the original state does not increase by cloning due to this noise. 
Recall that the inevitable noise in PIA comes from the same reason. 

Indeed, the optimal cloner can be constructed from the optimal phase-insensitive amplifier and beamsplitters. 
For example, $1\to2$ cloning with arbitrary asymmetry is achieved by putting an amplifier in one of the arms of a Mach-Zehnder interferometer~\cite{Fiurasek(2001):PRL}. 
Especially, for symmetric cloning, the reflectivity of the first beamsplitter becomes unity, i.e., it is achieved by first amplifying the original with $G=2$ and then splitting the amplified signal in half. 
This procedure can be extended to ${K}\to{L}$ cloning~\cite{Braunstein(2001):PRL,Fiurasek(2007):PRA}. 
The optimality of this realization is proven with respect to the cost function in Eq.~\eqref{eq:ClnCostFunc}~\cite{Fiurasek(2007):PRA}. 

For Gaussian cloning of coherent states, the added noise variance $n_k$ and the fidelity $F_k$ have correspondence as $F_k=1/(1+n_k)$. 
By using Eq.~\eqref{eq:ClnNoiseIneq}, the upper limit of fidelity is obtained for arbitrarily asymmetric Gaussian cloning. 
In particular, it becomes $F=2/3$ for the symmetric case. 
This is significantly higher than the classical limit of $F=1/2$, where we regard the limit of state estimation as the classical limit of symmetric cloning because the estimated state is classical information which can be copied any number of times. 
Note that the sameness of state estimation and asymptotic cloning where the number of clones tends to infinity is proven for a general set $S$ of possible original states~\cite{Bae(2006):PRL}. 
We refer to these fidelities only for the consistency with previous works. 
We stress again that our actual interest is the variances. 

We have seen above that the clones are made of the signal output of the amplifier. 
When cloning is unitarily realized, we still have the idler output, whose state is affected by the original state. 
Now we pay attention to this ancilla output system. 

As mentioned at the beginning of this section, anticlones are byproducts of cloning which are obtained from the ancilla systems. 
Especially, for $1\to2$ cloning, the idler output itself is an anticlone. 
It is an approximation of the phase-conjugated original state, or in other words, the output of an approximate NOT gate. 
Qubit version of this gate is demonstrated in Ref.~\cite{DeMartini(2002):Nature}. 

Anticlones are important when we are concerned with the reversibility. 
The originals can in principle be perfectly reproduced only when all the clones and anticlones are present. 
For the reversibility, the essential resource is nonclassical correlations, or entanglement. 
Conceptually, the excess noises in cloning is canceled by using the nonclassical correlations. 
Therefore, we discuss the existence of entanglement in the three-mode output system of $1\to2$ cloning. 
Clones are obtained by splitting the amplified signal, thus there is no entanglement among clones. 
However, there is entanglement between each clone and the anticlone. 
We stress that the resource for the recovery is not the anticlones themselves but the entanglement. 
We can obtain anticlones without entanglement with clones as follows. 
Suppose the situation where two independent cloners are running, and the same states are used as the inputs for them. 
Then, the clones obtained from one cloner do not have entanglement with the anticlones from the other. 
In this case, the originals can not be recovered from the noncorrelated outputs. 

For the recovery of the originals, the inverse unitary operation is not required. 
The optimal cloning can be fully reversed by the Bell measurement on a clone and an anticlone and subsequent feedforward to the remaining single clone~\cite{Filip(2004):PRA}. 
Note that this recovery scheme works not only for coherent states but also for an arbitrary core state $\ket{\psi}$. 
This scheme is efficient from two points of view. 
One is on a technical level that the homodyne measurements and feedforward displacement operations are quite efficient with current technology. 
The other is on a conceptual level that the performer of the Bell measurement and the owner of the remaining clone who is willing to recover the original can be spatially separated. 
For this case, they only need classical channels for communication, and never quantum channels. 
Note that even partial reversal is possible with a similar scheme based on local operations and classical communication (LOCC), which converts, e.g., symmetric clones to asymmetric clones~\cite{Filip(2004):PRA}. 

We would like to discuss practical aspects of cloning and its reversibility assisted by classical communication. 
As described above, cloning of a quantum state is regarded as distribution of information among plural participants. 
The information of the original is to some extend accessible to individual participants. 
This situation is clearly distinguished from that found in usual quantum error correcting codes where the quantum information is mapped on a larger Hilbert space so that no information about the original is accessible from a localized system. 
Such share of information would play important roles in several scenarios, in which the reversibility would give a tactical aspect to information exchange. 
For example, cloning is a possible attack by an eavesdropper in QKD. 
In this example, the reversibility of cloning provides the opportunity for the communicators to negotiate with the eavesdropper when they know the attack~\cite{Filip(2004):PRA}. 
Since coherent states are a strong candidate for the information carrier in quantum communication, cloning of coherent states is especially of great significance. 

There are several experimental previous works which demonstrate cloning of coherent states beyond the classical limit of $F=1/2$ in non-reversible ways, i.e., their anticlones are lost in the environment. 
In Ref.~\cite{Andersen(2005):PRL}, using feedforward-based PIA of Ref.~\cite{Josse(2006):PRL}, almost quantum-limited $1\to2$ cloning is demonstrated. 
In Ref.~\cite{Koike(2006):PRL}, telecloning is demonstrated where the original coherent state is teleported and cloned at the same time. 

In Sec.~\ref{sec:ResultsClone}, we demonstrate $1\to2$ symmetric Gaussian cloner which preserves an anticlone at the output. 
As is shown in Fig.~\ref{sfig:ClnSchematic}, we apply a half beamsplitting to the signal output of the feedforward-based PIA with the gain $G=2$ described in Sec.~\ref{sec:FfPia}. 
In the demonstration, the reversibility is checked from the output correlations. 
To our knowledge, there is no previous experiment of this kind even in qubit regime. 
Although our demonstration is only for coherent states, our cloner should equally work for arbitrary core state $\ket{\psi}$ as discussed above. 

We close this section by giving the input-output relation of the optimal $1\to2$ symmetric cloning. 
By substituting $G=2$ and $\theta=0$ into the input-output relation in Eq.~\eqref{eq:PiaUnitary} and splitting the signal output in half, we obtain, 
\begin{subequations}
\begin{align}
\hat{a}_\text{cln-1}= &
\hat{a}_\text{org}+\tfrac{1}{\sqrt{2}}\hatd{a}_\text{idl}+\tfrac{1}{\sqrt{2}}\hat{a}_\text{vac}, \\
\hat{a}_\text{cln-2}= &
\hat{a}_\text{org}+\tfrac{1}{\sqrt{2}}\hatd{a}_\text{idl}-\tfrac{1}{\sqrt{2}}\hat{a}_\text{vac}, \\
\hat{a}_\text{a-cln}= &
\hatd{a}_\text{org}+\sqrt{2}\,\hat{a}_\text{idl},
\end{align}
\end{subequations}
where the subscripts `org', `\mbox{cln-1}', `\mbox{cln-2}', and `\mbox{a-cln}' denote the original, first clone, second clone, and anticlone, respectively. 
The subscript `idl' denotes the idler input for PIA which is in a vacuum state. 
The annihilation operators with the subscript `vac' indicate another ancilla in a vacuum state, which invades from the empty port of the final half beamsplitter. 
For the excess noise of the two clones, $n_1=n_2=1/2$ is easily checked. 
Therefore, this cloner is optimum when evaluated by the cost function in Eq.~\eqref{eq:ClnCostFunc} with $c_1=c_2$. 
When PIA is realized with a feedforward-based scheme as is found in Sec.~\ref{sec:ResultsClone}, further excess noise contaminates the output in accordance with the squeezing levels of the ancillas.

\section{Experimental Setup}
\label{sec:SetUp}

\begin{figure*}[!tb]
\centering
\subfigure[Setup for PIA.]{
\includegraphics[clip,scale=0.58]{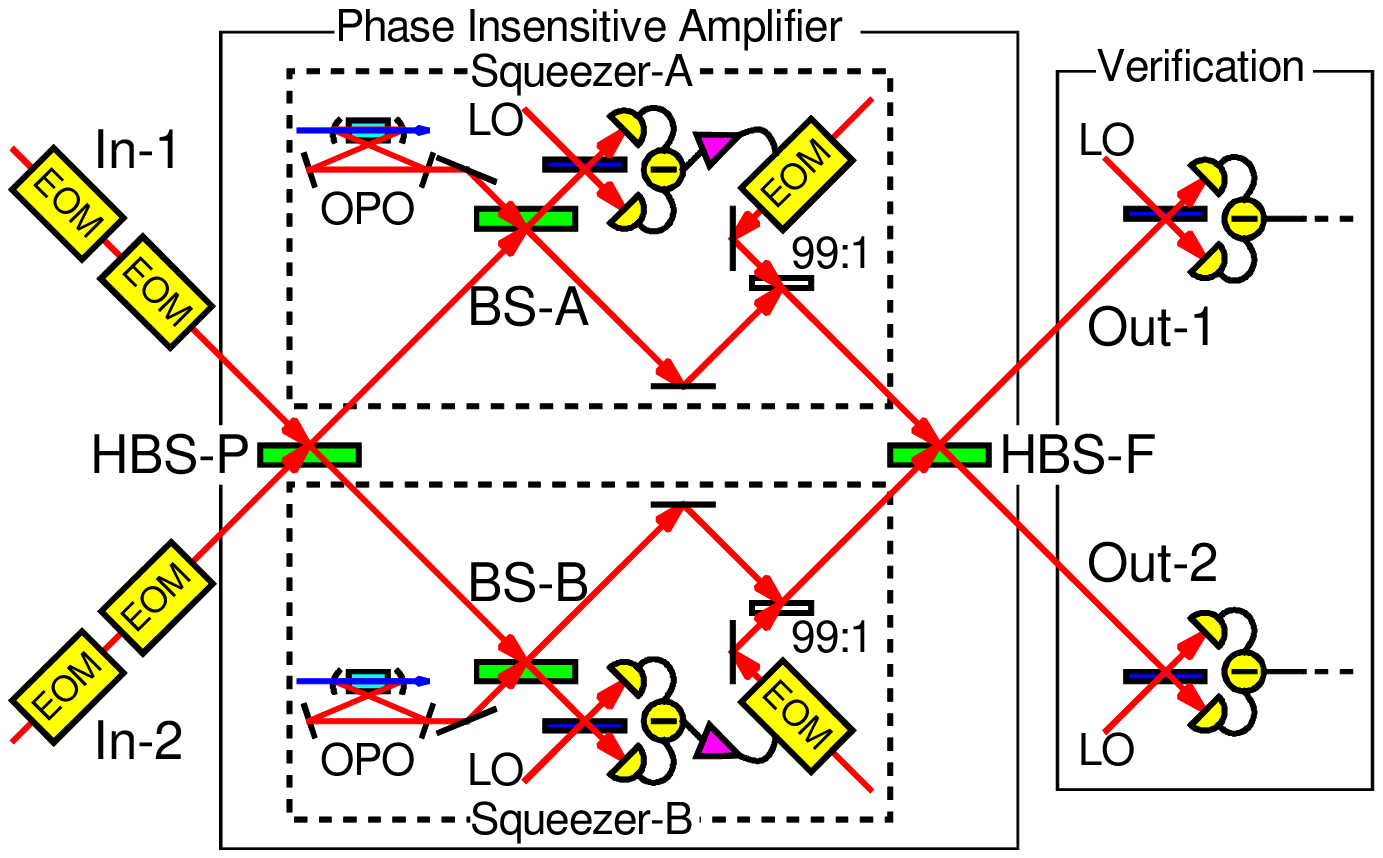}
\label{sfig:AmpSchematic}
}
\hfill
\subfigure[Setup for $1{\to}2$ cloner.]{
\includegraphics[clip,scale=0.58]{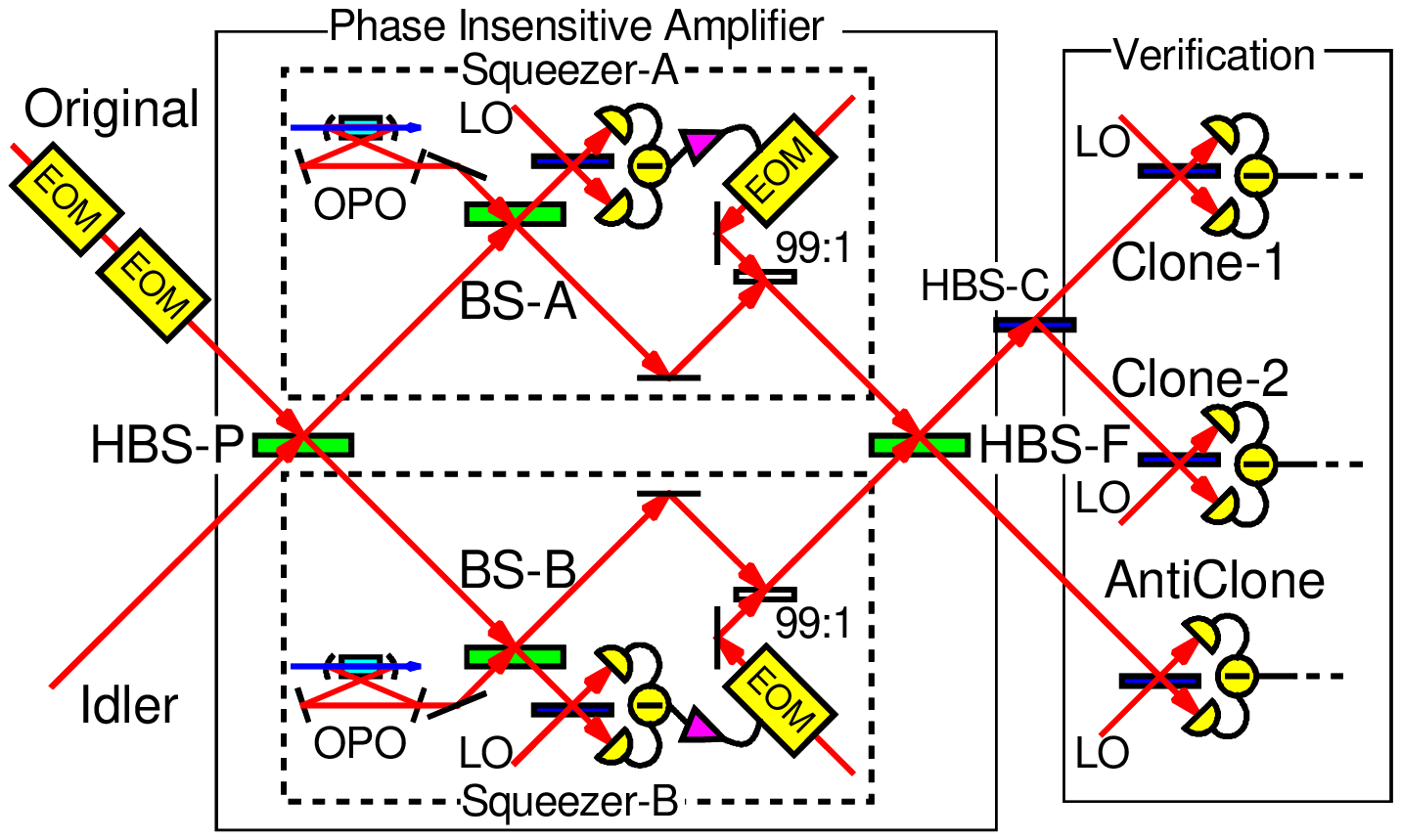}
\label{sfig:ClnSchematic}
}
\caption{
Schematics of experimental setups.
OPO:~Optical parametric oscillator. 
EOM:~Electro-optic modulator. 
LO:~Local oscillator. 
\mbox{HBS-P}, \mbox{HBS-F}, \mbox{HBS-C}:~Half beamsplitter.
\mbox{BS-A}, \mbox{BS-B}:~Beamplistter with reflectivity $R\approx0.17$.
Four beamsplitters (\mbox{HBS-P}, \mbox{HBS-F}, \mbox{BS-A}, and \mbox{BS-B}) are variable, composed of two polarization beamsplitters and one half-wave plate.
}
\label{fig:Schematic}
\end{figure*}

Schematic of the experimental setup for PIA is illustrated in Fig.~\ref{sfig:AmpSchematic}, and that for approximate cloning is illustrated in Fig.~\ref{sfig:ClnSchematic}. 
The light source is a Ti:sapphire laser, which has a continuous-wave single-mode output of $860$~nm in wavelength and about $1.5$~W in power.
We treat the quantum states of narrow sidebands located at $1.34$~MHz apart from the optical carrier frequency. 

Two main beams that go from \mbox{in-1} and \mbox{in-2} to \mbox{out-1} and \mbox{out-2} in Fig.~\ref{sfig:AmpSchematic} carry the quantum states which are targets of PIA. 
The setup has a form of a Mach-Zehnder interferometer that holds a single-mode squeezer (\mbox{squeezer-A} or \mbox{squeezer-B}) in each arm. 
This decomposition of unitary PIA into squeezers and beamsplitters is derived from the bosonic version of Bloch-Messiah reduction shown in Ref.~\cite{Braunstein(2005):PRA}. 
We note that this setup is almost the same as that for quantum nondemolition (QND) interaction demonstrated in Ref.~\cite{Yoshikawa(2008):PRL}. 
This fact shows the capability of our setup to realize many types of two-mode Gaussian interaction. 
Combining PIA for $G=2.0$ with another half beamsplitter as is shown in Fig.~\ref{sfig:ClnSchematic}, $1\to2$ approximate cloning of coherent states is achieved. 

\mbox{Squeezer-A} and \mbox{squeezer-B} are feedforward-based squeezers, which are theoretically proposed in Ref.~\cite{Filip(2005):PRA} and experimentally demonstrated in Ref.~\cite{Yoshikawa(2007):PRA}. 
Each squeezer consumes an ancilla in a squeezed state, which is generated by an optical parametric oscillator (OPO). 

Note that several essential optical elements are omitted from Fig.~\ref{fig:Schematic}, such as a second harmonic generation (SHG) cavity to generate pump beams for OPOs, and three spatial-mode cleaning cavities (MCCs). 
One MCC is used for local oscillators (LOs) for homodyne measurements and auxiliary beams for feedforward displacements. The other two MCCs are used for individual input beams.

The experimental procedure is divided into three steps: 
Firstly, we prepare input coherent states and ancilla squeezed vacuum states. 
Secondly, we implement PIA and cloning via feedforward. 
Finally, the output states are homodyne measured for verification.
In the following, we describe the experimental details of each step.

\subsection{Preparation}

At this step, we generate coherent states which are used as inputs, and squeezed vacuum states which are used as ancillas.

The nonzero mean values of the sideband coherent states at $1.34$~MHz are produced by appropriately modulating the optical carriers. 
In our setup, the relative phase of interference at each beamsplitter is designed to be fixed with active feedback control. 
Therefore, in order to make an arbitrary phase space displacement in the input modes, both amplitude modulation (AM) and phase modulation (PM) are utilized. 
AM and PM make non-zero mean values of $\hat{x}_1^\text{in}$ and $\hat{p}_1^\text{in}$ for the first input mode, and those of $\hat{p}_2^\text{in}$ and $\hat{x}_2^\text{in}$ for the second input mode, respectively. 
Each of four electro-optic modulators (EOMs) before PIA in Fig.~\ref{sfig:AmpSchematic} corresponds to one of these four quadratures. 
On the other hand, in Fig.~\ref{sfig:ClnSchematic}, only two EOMs are depicted before PIA which are both located at the first input beam path. 
Therefore, the symmetry of the two input modes is broken in the cloning experiment.
One input mode is the target of cloning, and the other input mode is set in a vacuum state throughout. 
For both experiments, the modulations are switched on and off in order to use several coherent states as inputs. 
After these EOMs, there are MCCs though they are omitted from Fig.~\ref{fig:Schematic}.

Squeezed vacuum states are each generated by an OPO which is driven below the threshold. 
Our OPO has a bow-tie shaped configuration with a round-trip length of about 500~mm. 
It contains a periodically-poled KTiOPO$_4$ (PPKTP) crystal as a nonlinear optical medium, which is commercially available from Raicol and has 10~mm length and 1~mm by 1~mm cross section.
The experimental datails of our OPO squeezing are found in Ref.~\cite{Suzuki(2006):APL}.
The squeezing level with the pump of about $100$~mW is about $-5$~dB relative to the shot noise level at $1.34$~MHz. 
The pump beams for the OPOs are the second harmonic of a fundamental beam, generated by a SHG cavity. 
Most of the Ti:sapphire laser output is sent to the SHG cavity, whose output of about $300$~mW is divided into two to pump the individual OPOs.
The SHG cavity has almost the same configuration as that of OPOs, whereas a KNbO$_3$ (KN) crystal is used instead of the PPKTP crystal. 

Modulation sidebands other than $1.34$~MHz are exploited for active feedback control of optical interferences.
A modulation at $13.5$~MHz is utilized for locking cavities, including the SHG cavity, the two OPOs, and the three MCCs. 
On the other hand, lower frequency modulations at $193$~kHz and $333$~kHz are utilized at the OPOs to lock the phases of the pump beams. 
Furthermore, the two input beams are modulated at $108$~kHz and $154$~kHz. 
These four low-frequency modulations contribute to the lock of the downstream interferometric system in the subsequent steps, as mentioned later.

\subsection{Amplifier and Cloner}

The two input beams are combined at a preceding half beamsplitter (\mbox{HBS-P}) and then sent to \mbox{squeezer-A} and \mbox{squeezer-B}. 
After the squeezing operations, the two beams interfere again at another half beamsplitter (\mbox{HBS-F}), which completes PIA. 
By splitting one of the two output beams by another half beamsplitter (\mbox{HBS-C}), $1\to2$ cloner is obtained. 

The squeezing procedure goes as follows. 
First, the main beam is combined with an ancilla beam coming from an OPO at a beamsplitter (\mbox{BS-A} or \mbox{BS-B}). 
Next, one of the two beams after the beamsplitter is homodyne measured. 
Finally, the measurement outcome is fed forward to the remaining beam. 
\mbox{BS-A} and \mbox{BS-B} have the common reflectivity of $R$.
This parameter $R$ determines the degree of the feedforward-based squeezing and thus the gain of amplification $G$ with the relation shown in Eq.~\eqref{eq:Gain}. 
As is already mentioned, $R\approx0.17$ for our demonstration of $G=2.0$.

The feedforward operation is a phase space displacement whose amount is proportional to the random outcome of the homodyne measurement.
The electric signal from the homodyne detector is sent to an EOM to be converted into an optical signal, where the gain and phase at $1.34$~MHz are carefully chosen.
The auxiliary beam which is modulated by the feedforward EOM has the power of $150$~$\mu$W, $1\%$ of which subsequently enters the mainstream via an asymmetric beamsplitter (99:1).

The powers of the two input beams are $10$~$\mu$W, and those of the two ancilla beams are $2$~$\mu$W. 
These powers are considerably smaller than $3$~mW of LOs used for homodyne detections.

The four beamsplitters of PIA (\mbox{HBS-P}, \mbox{HBS-F}, \mbox{BS-A}, and \mbox{BS-B}) are actually composed of two polarization beam splitters and a half-wave plate in the same manner as the QND experiment in Ref.~\cite{Yoshikawa(2008):PRL}. 
Their reflectivities are variable by rotating half-wave plates. 
They enable us to measure the input states as well as the output states with the same homodyne detectors for verification.
The propagation losses of two main beams are measured to be $7\%$ on average, which mostly come from these variable beam splitters. 

In order to control the relative phases at beamsplitters with active feedback, interferences between the carriers and the low-frequency modulations are monitored. 
This is typically done by picking up $1\%$ of the beam after the interference, though such details are omitted from Fig.~\ref{fig:Schematic}. 
For each locking point, an appropriate modulation sideband is chosen, and the error signal is extracted from the interference between the carrier and the sideband by demodulation.
However, \mbox{HBS-P} and \mbox{HBS-F} are exceptions, where the interference between two modulation sidebands, namely $108$~kHz and $154$~kHz, is exploited. 
The beat frequency of $46$~kHz is chosen for the reference signal of demodulation to obtain the error signals.

\subsection{Verification}

PIA is characterized by measuring two-mode input states as well as two-mode output states using two homodyne detections.
In the cloning experiment, on the other hand, three-mode output states are compared with single-mode input states. 
The input states are measured by setting the reflectivities of the four variable beamsplitters to unity and disabling the feedforward. 
The quantum efficiency of a homodyne detector is about $99\%$, and the dark noise is about $17$~dB below the optical shot noise produced by the LO. 
The interference visibilities to the LOs are $98\%$ on average.

The outcomes of the final homodyne measurements are analyzed in either of the two ways below.

In one way of analysis, the quadrature data are directly treated, which are obtained by lock-in detection of $1.34$~MHz components of the homodyne outputs. 
A signal from a homodyne detector is mixed with the reference signal at $1.34$~MHz, and then low-pass filtered with the cutoff of $30$~kHz. 
Subsequently it is analog-to-digital (A/D) converted for storage with the sampling rate of $300$~kHz and the resolution of $14$ bits (\mbox{PXI-5122}, National Instruments Corporation).
In this analysis, the phase of the homodyne detection is slowly scanned. 
The phase information is stored simultaneously with the quadrature values using the same A/D board.
From the resulting marginal distributions, phase space distributions (i.e., Wigner functions) are reconstructed, where we assume that all the quantum states obtained in the experiments are Gaussian.
The first and second moments are computed so that the likelihoods are maximized.

The other way is the power analysis at $1.34$~MHz using a spectrum analyzer. 
In this analysis, the measured quadratures are set to either $\hat{x}$ or $\hat{p}$.
Not only the powers of the output quadratures but also those of their correlations are measured for several input coherent states.
The resolution bandwidth is $30$~kHz, the video bandwidth is $300$~Hz, the sweep time is $0.1$~s, and $20$ times averaging is taken for each trace. 

Note that we can easily see the effect of the Hermitian conjugate term in Eq.~\eqref{seq:PiaIdler} as a mirror image with the former way of analysis, whereas we can not do this with the latter.

\section{Experimental Results for Phase-Insensitive Amplifier}
\label{sec:ResultsPia}

The two main modes are denoted by ``\mbox{mode-1}'' and ``\mbox{mode-2}''.
One of them is the ``signal'' and the other is the ``idler'', which are initially in a coherent state and a vacuum state, respectively.
By swapping the role of the signal and idler, we check the symmetry of our PIA.

\begin{figure}[!t]
\centering
\subfigure[Signal input.]{
\includegraphics[clip,scale=0.319]{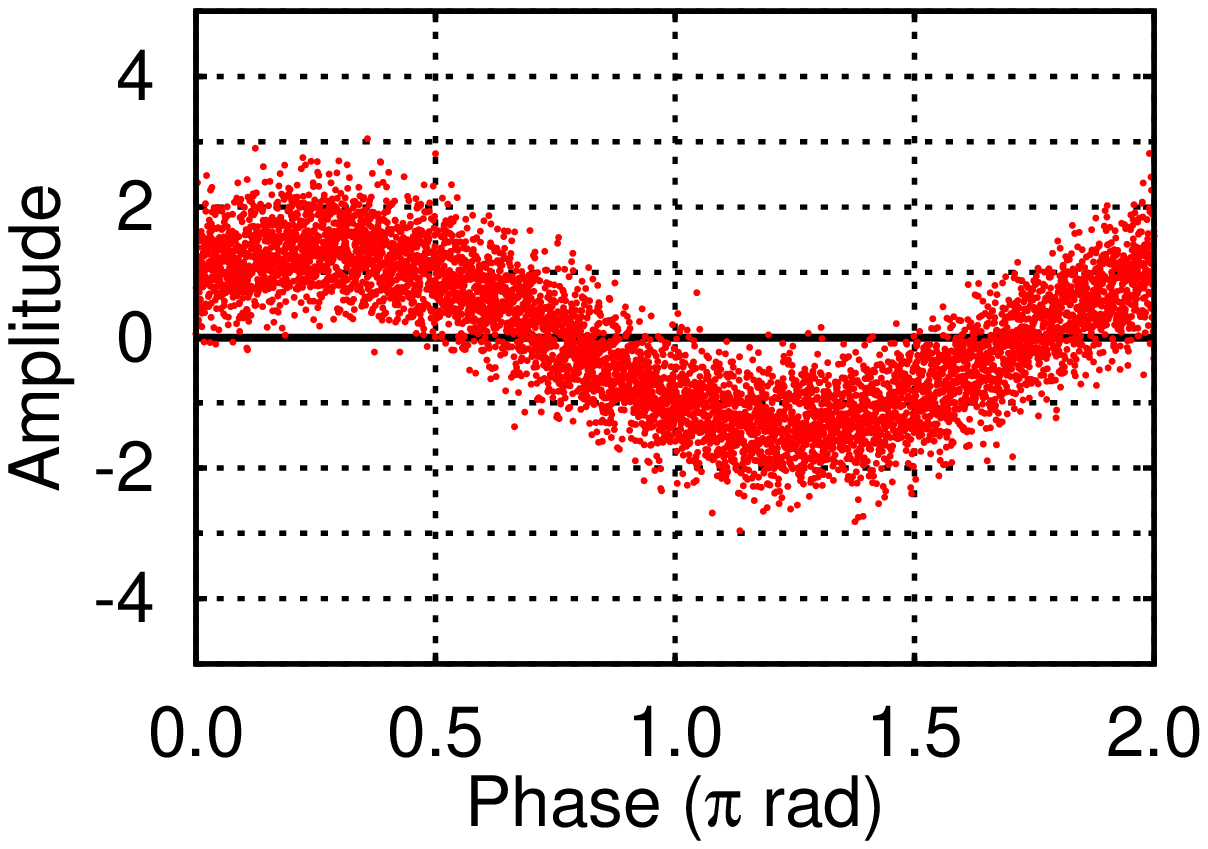}
\label{sfig:Amp_SigIn1}
}\\
\subfigure[Signal output.]{
\includegraphics[clip,scale=0.319]{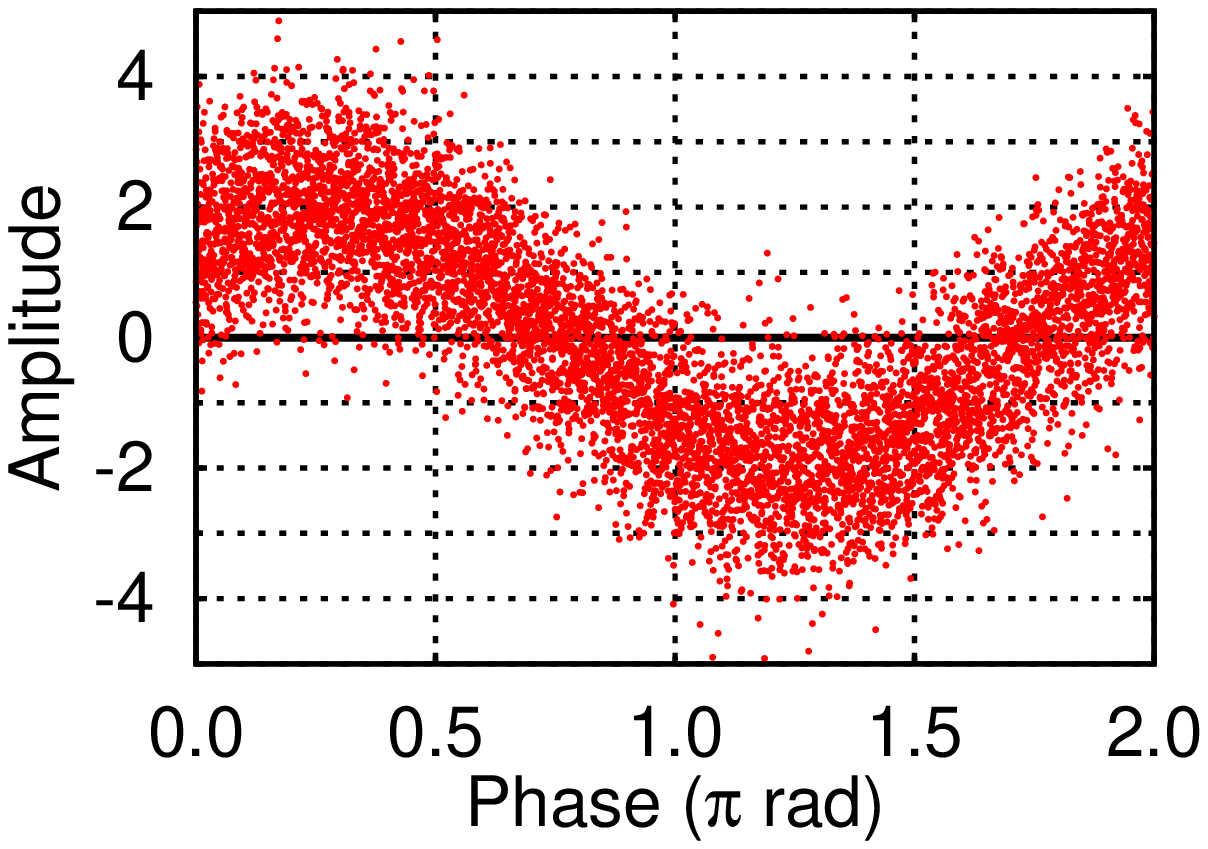}
\label{sfig:Amp_SigOut1}
}
\subfigure[Idler output.]{
\includegraphics[clip,scale=0.319]{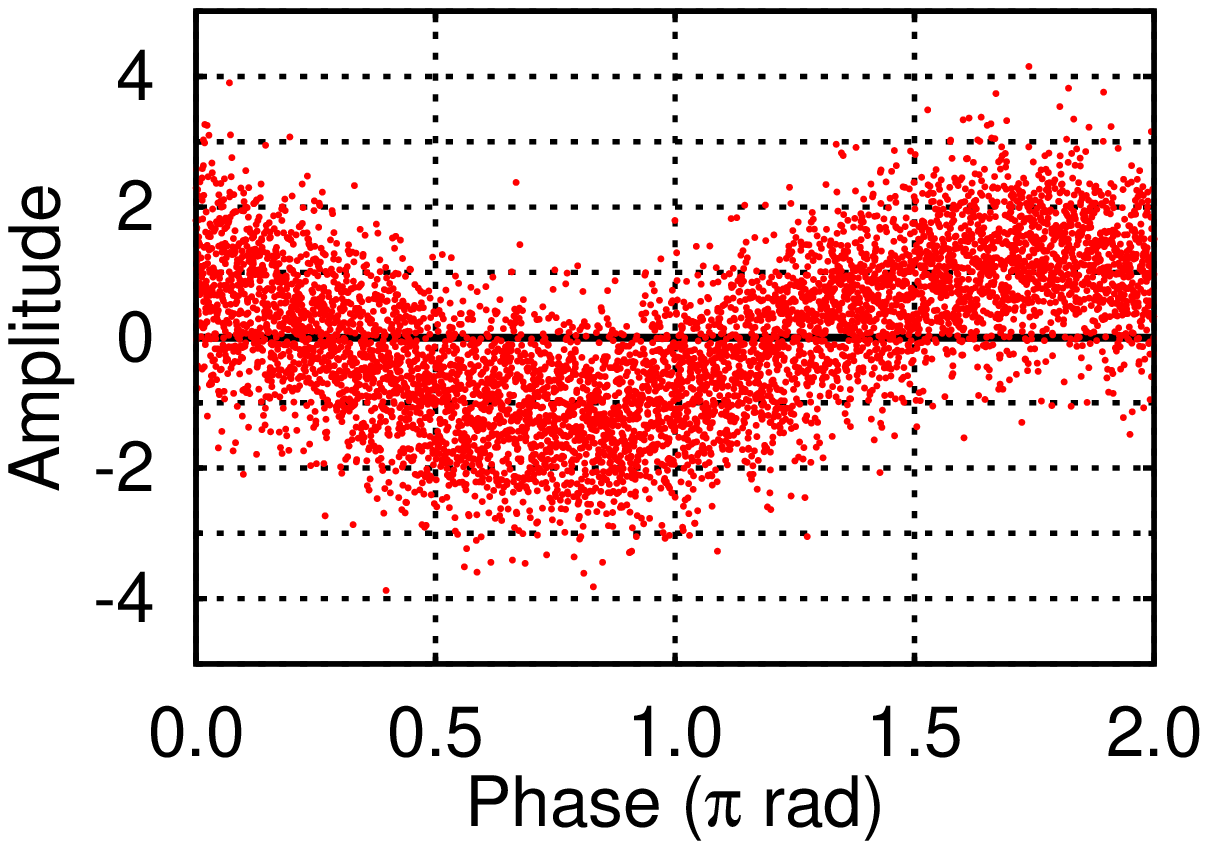}
\label{sfig:Amp_IdlerOut1}
}
\caption{
Quadrature data. 
\mbox{Mode-1} for signal and \mbox{mode-2} for idler.
The phases of homodyne measurements are scanned from $0$ to $2\pi$, which correspond to horizontal axes.
Vertical axes are quadrature values.
}\label{fig:Amp_MD1}
\subfigure[Experiment.]{
\includegraphics[clip,scale=0.45]{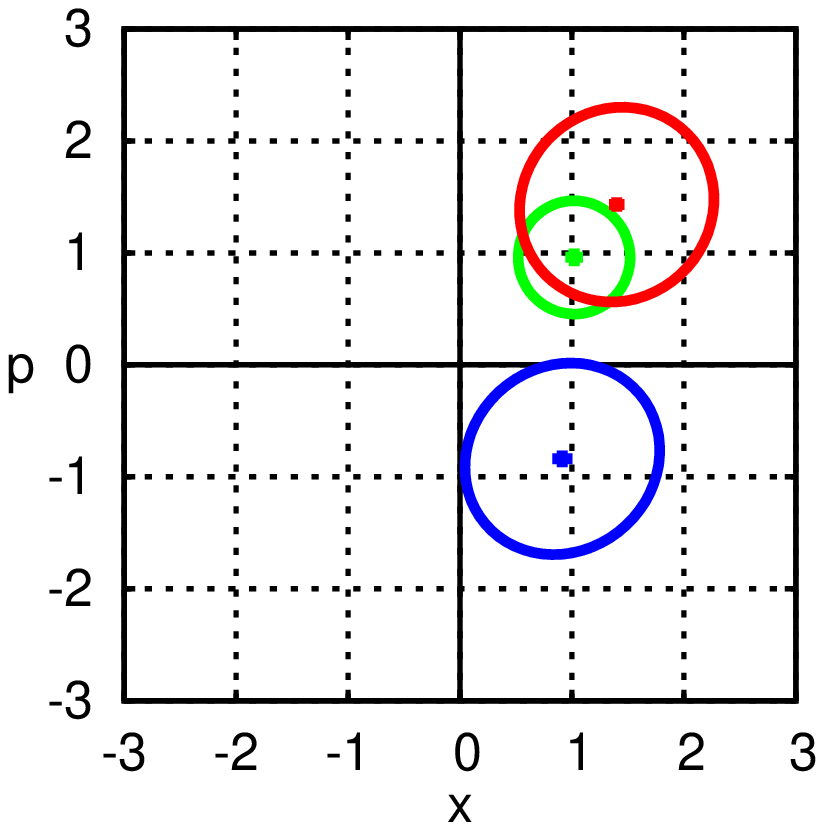}
}
\subfigure[Theory.]{
\includegraphics[clip,scale=0.45]{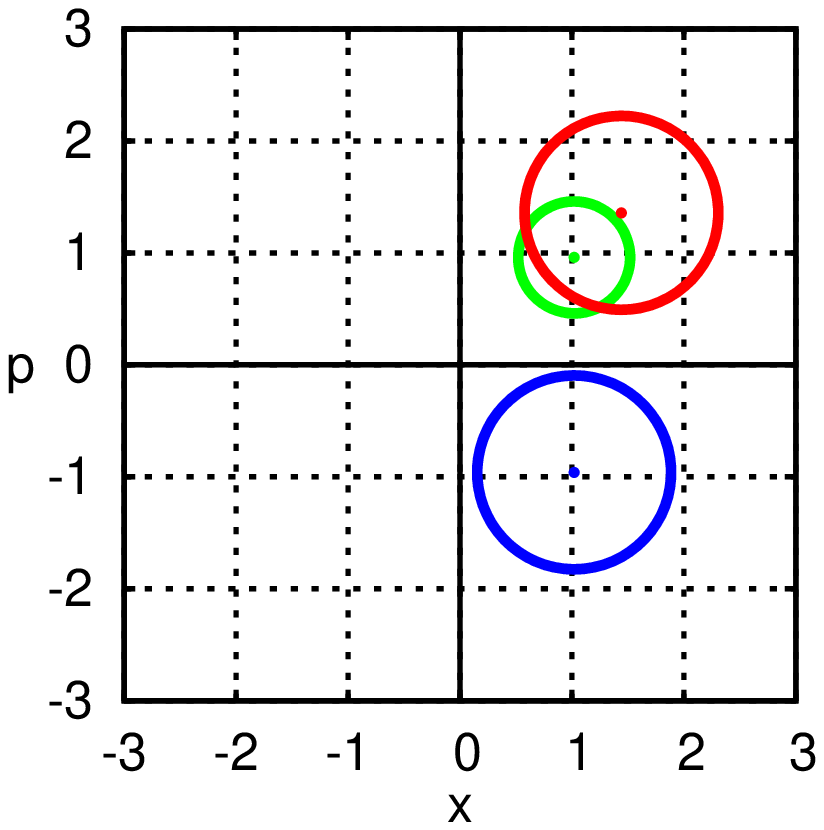}
}
\caption{
Phase space distributions computed from the quadrature data in Fig.~\ref{fig:Amp_MD1}.
The first and second moments of Gaussian Wigner functions are represented by ellipses.
Green:~Signal input.
Red:~Signal output.
Blue:~Idler output.
}\label{fig:Amp_PS1}
\end{figure}

\begin{figure}[!t]
\centering
\subfigure[Signal input.]{
\includegraphics[clip,scale=0.319]{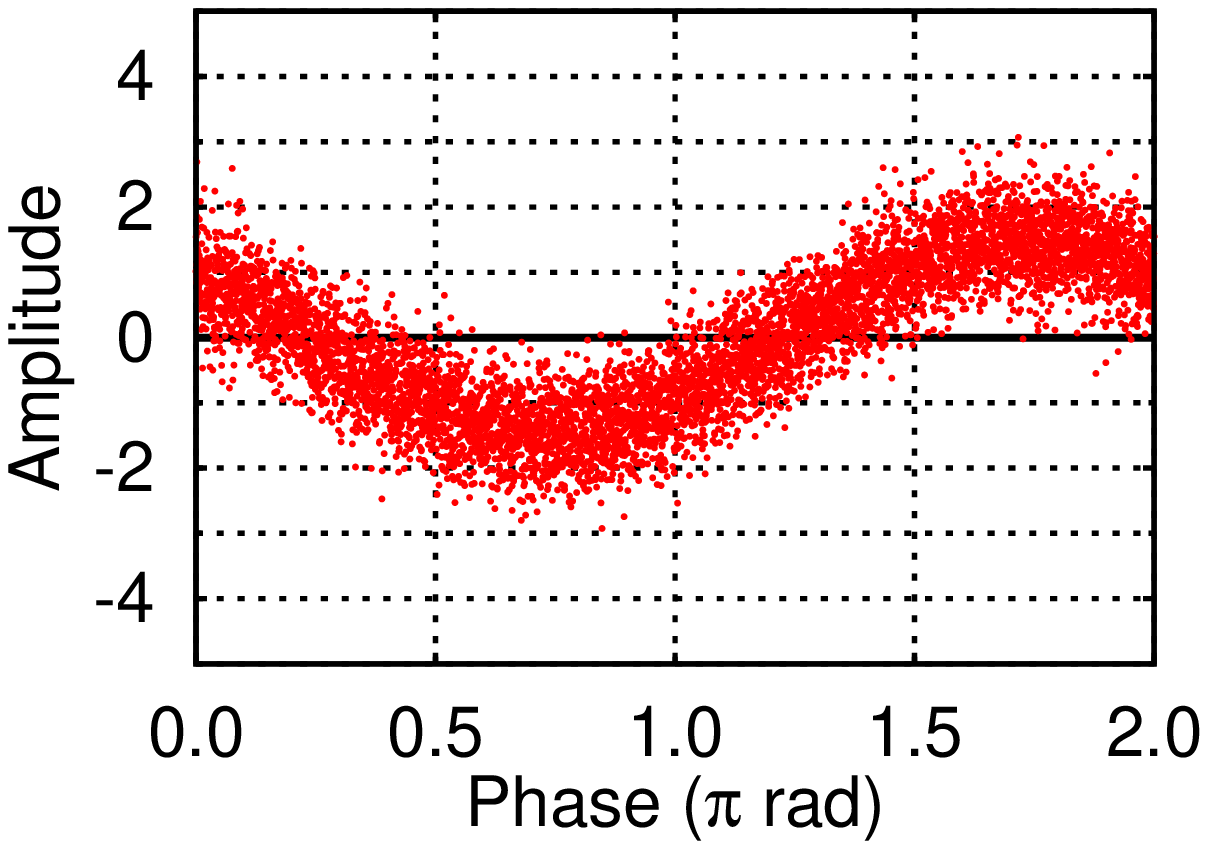}
\label{sfig:Amp_SigIn2}
}\\
\subfigure[Signal output.]{
\includegraphics[clip,scale=0.319]{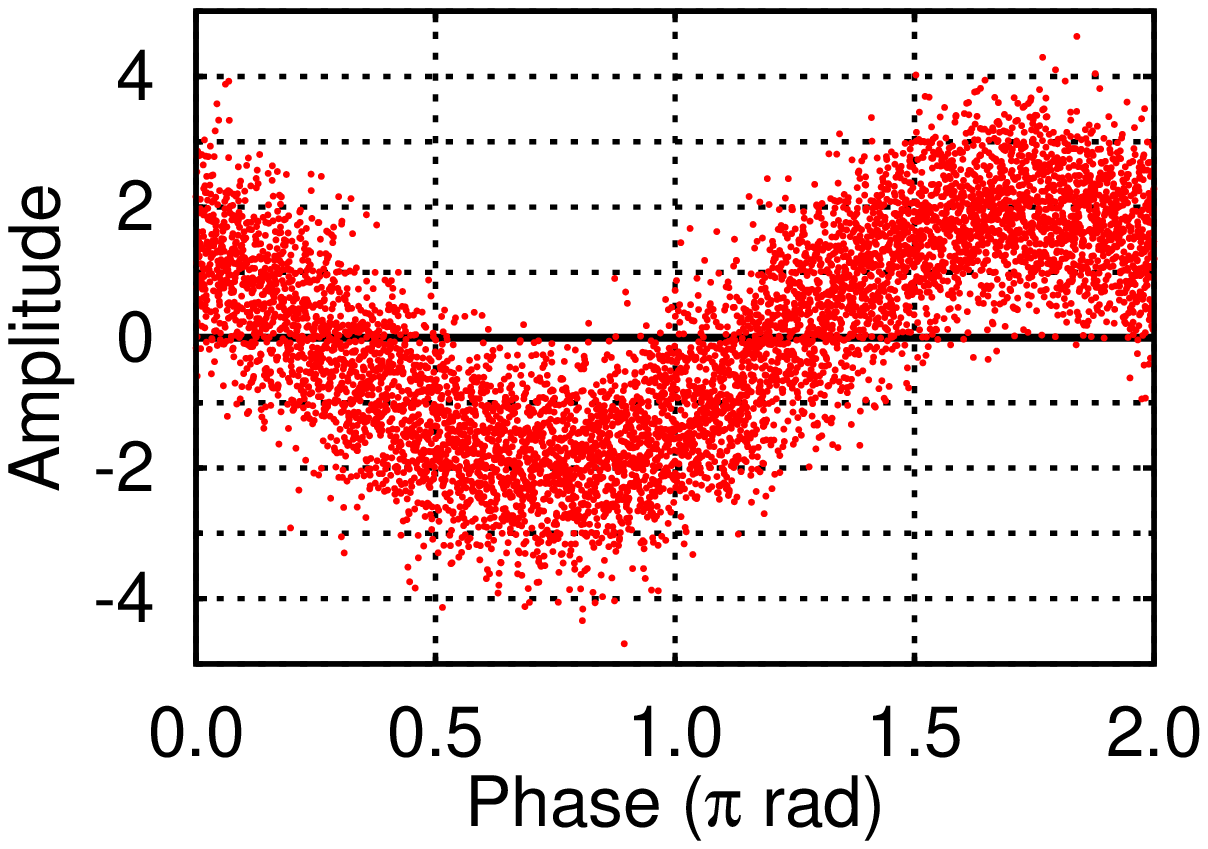}
\label{sfig:Amp_SigOut2}
}
\subfigure[Idler output.]{
\includegraphics[clip,scale=0.319]{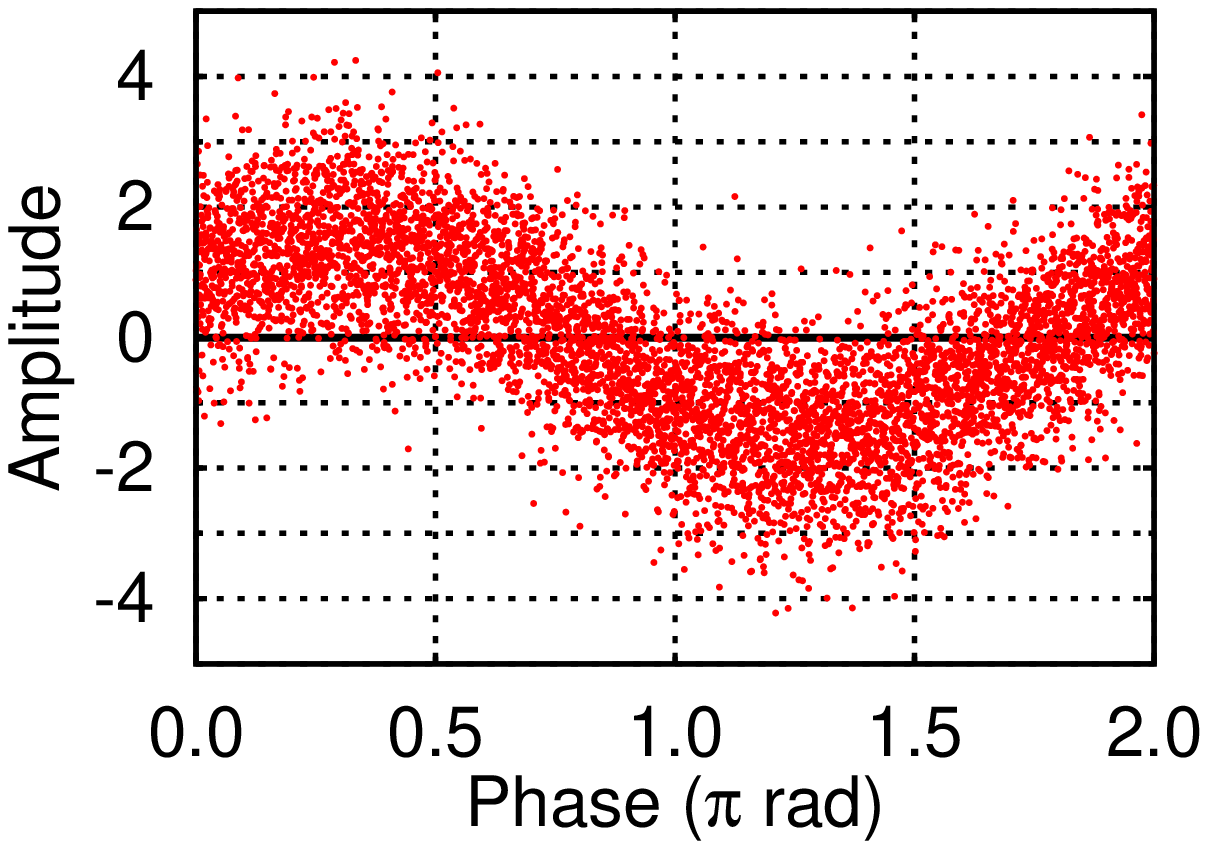}
\label{sfig:Amp_IdlerOut2}
}
\caption{
Quadrature data. 
\mbox{Mode-2} for signal and \mbox{mode-1} for idler.
The phases of homodyne measurements are scanned from $0$ to $2\pi$, which correspond to horizontal axes.
Vertical axes are quadrature values.
}\label{fig:Amp_MD2}
\subfigure[Experiment.]{
\includegraphics[clip,scale=0.45]{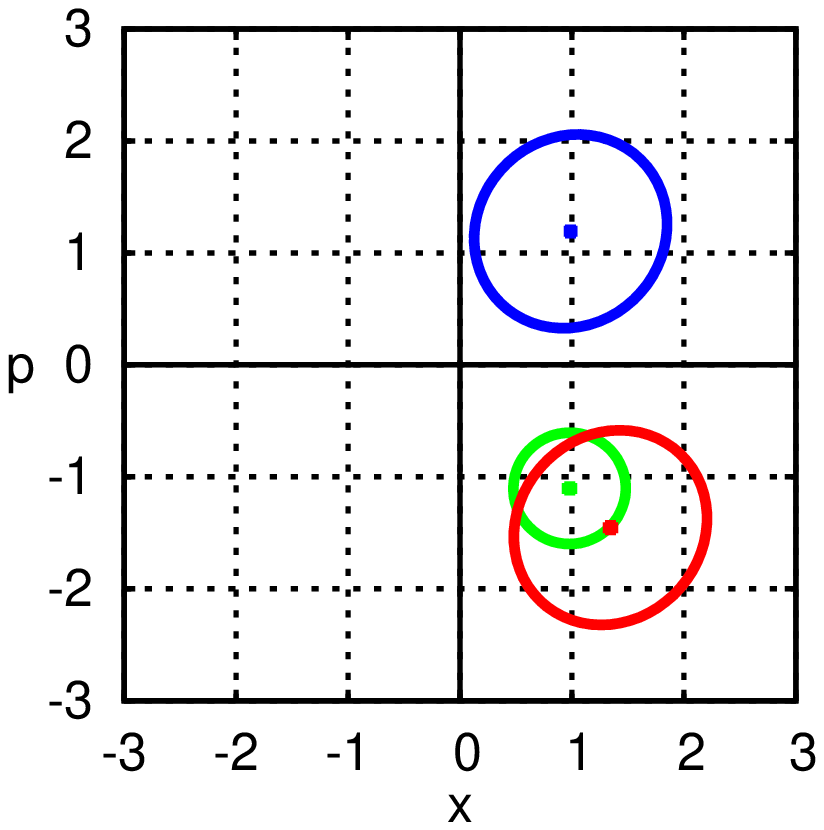}
}
\subfigure[Theory.]{
\includegraphics[clip,scale=0.45]{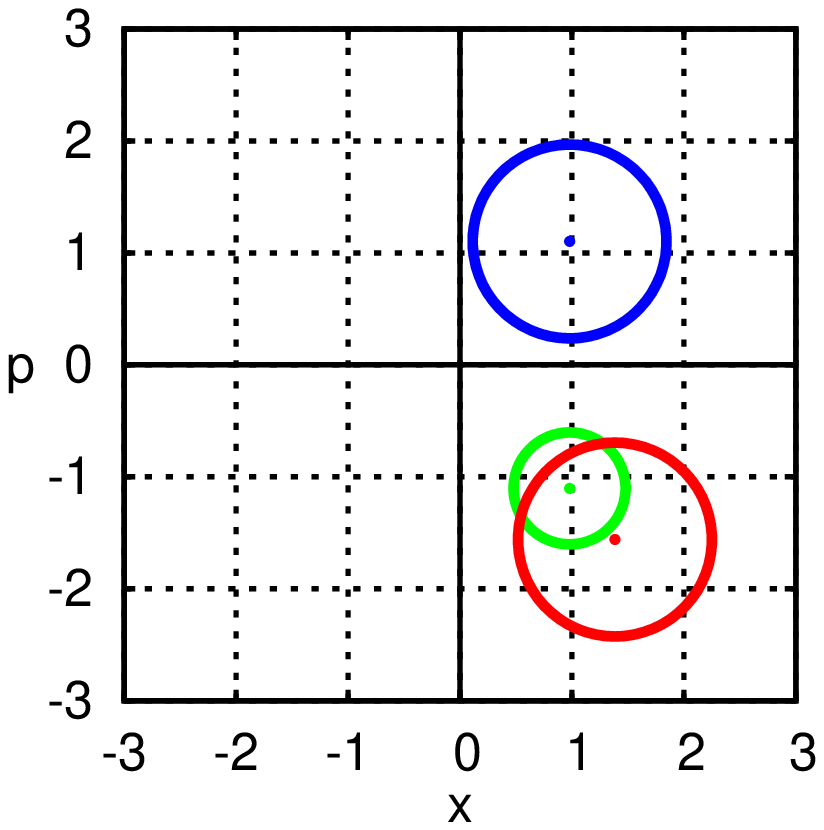}
}
\caption{
Phase space distributions computed from the quadrature data in Fig.~\ref{fig:Amp_MD2}.
The first and second moments of Gaussian Wigner functions are represented by ellipses.
Green:~Signal input.
Red:~Signal output.
Blue:~Idler output.
}\label{fig:Amp_PS2}
\end{figure}

We first show the results of the lock-in detection, because it is intuitively easier to see. 
Figs.~\ref{fig:Amp_MD1} and \ref{fig:Amp_MD2} show the experimental quadrature values at various phases of LOs. 
For Fig.~\ref{fig:Amp_MD1}, the \mbox{mode-1} is the signal and the \mbox{mode-2} is the idler, whereas for Fig.~\ref{fig:Amp_MD2}, the \mbox{mode-2} is the signal and the \mbox{mode-1} is the idler. 
There are three subfigures corresponging to the signal input~(a), the signal output~(b), and the idler output~(c). 
Horizontal axes are the measurement phases $\phi$ which are scanned from $0$ to $2\pi$.
The quadrature at $\phi=0$ corresponds to $\hat{x}$ and that at $\phi=\pi/2$ does to $\hat{p}$.
Vertical axes are normalized quadrature values where the standard deviation of vacuum fluctuation is $0.5$. 
Each set of data is taken for about $0.2$ seconds. 
Quadrature data are plotted every $10$ points in the figures, whereas the whole data are used for the analysis. 
The sinusoidal curve of the signal input represents the nonzero mean amplitude of a coherent state, and the fluctuation around the sinusoid represents the quantum noise. 
With regard to the fluctuation, it grows uniformly at both the signal and idler outputs. 
This uniformity is an evidence of the phase-insensitivity of our amplifier. 
On the other hand, with regard to the sinusoidal curve, the two output modes show different behaviors. 
At the signal output, the amplitude of the sinusoid is amplified from that of the signal input, maintaining the phase. 
At the idler output, the amplitude of the sinusoid is the same as the signal input, whereas the phase is flipped. 
This flip is due to the Hermitian conjugate term in Eq.~\eqref{seq:PiaIdler}. 
For both figures, the same qualitative behaviors as mentioned above are observed. 

Figs.~\ref{fig:Amp_PS1} and \ref{fig:Amp_PS2} are phase space diagrams, which are computed from the quadrature data shown in Figs.~\ref{fig:Amp_MD1} and \ref{fig:Amp_MD2}, respectively.
The experimental results~(a) and the theoretical calculations for the optimal PIA~(b) are depicted next to each other. 
In the theoretical calculations, the experimental value is used for the amplitude of the signal input. 
The first and second moments are expressed by ellipses, which correspond to the cross sections of the Wigner functions. 
Note that the theoretical ellipses in~(b) are strictly circles. 
In each phase space diagram, there are three ellipses. 
The ellipse in green is the signal input. 
Its radius is almost $0.5$ which corresponds to the standard deviation of vacuum fluctuation.
The ellipse in red is the signal output. 
Its center is about $\sqrt{2}$ times farther away from the origin and its radius is about $\sqrt{3}$ times larger than those of the signal input. 
The ellipse in blue is the idler output. 
Its radius is almost the same as that of the signal output, whereas its center is flipped around the $x$-axes from that of the signal input, which again represents the Hermitian conjugate term in Eq.~\eqref{seq:PiaIdler}.

\begin{figure}[!t]
\centering
\includegraphics[clip,scale=0.478]{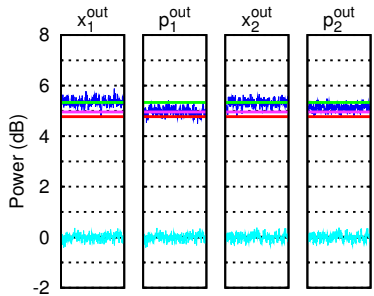}
\caption{
Output powers for vacuum inputs.
Vertical axes are powers in dB scale normalized by shot noises.
Blue:~Output quadratures.
Cyan:~Shot noises.
Red:~Theory for optimal PIA outputs.
Magenta:~Theory for our PIA outputs with $-5$~dB squeezed ancillas.
Green:~Theory for our PIA outputs with vacuum ancillas.
}\label{fig:Amp_V}
\end{figure}

\begin{figure}[!t]
\centering
\subfigure[$\avg{\hat{x}_1^\text{in}}\neq0$.]{
\includegraphics[clip,scale=0.478]{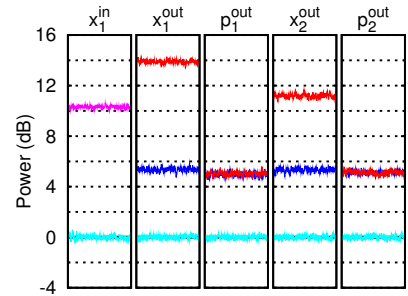}
}
\subfigure[$\avg{\hat{p}_1^\text{in}}\neq0$.]{
\includegraphics[clip,scale=0.478]{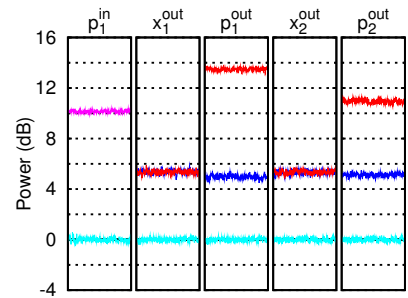}
}
\subfigure[$\avg{\hat{x}_2^\text{in}}\neq0$.]{
\includegraphics[clip,scale=0.478]{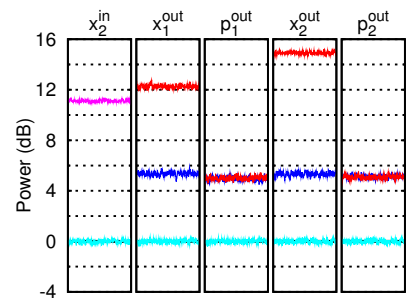}
}
\subfigure[$\avg{\hat{p}_2^\text{in}}\neq0$.]{
\includegraphics[clip,scale=0.478]{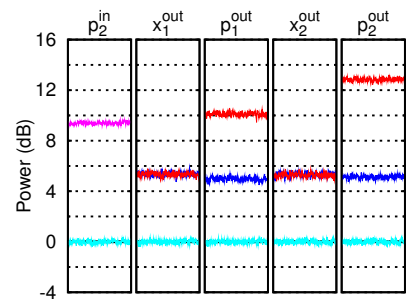}
}
\caption{
Output powers for inputs in several coherent states.
One of four input quadratures $\hat{x}_1^\text{in}$, $\hat{p}_1^\text{in}$, $\hat{x}_2^\text{in}$, $\hat{p}_2^\text{in}$ is excited, at the same time the other three quadratures are left at the vacuum level.
Vertical axes are powers in dB scale normalized by shot noises.
Magenta:~Excited input quadratures.
Red:~Output quadratures with excitation in inputs.
Blue:~Output quadratures without excitation in inputs.
Cyan:~Shot noises.
}\label{fig:Amp_C}
\end{figure}

\begin{figure}[!t]
\centering
\includegraphics[clip,scale=0.478]{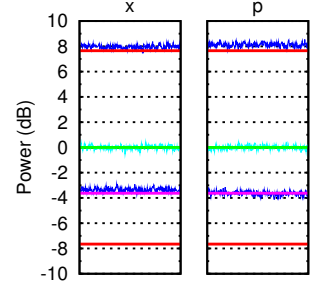}
\caption{
Two-mode squeezing and antisqueezing for vacuum inputs.
Vertical axes are powers in dB scale normalized by summed shot noises of two homodyne detections.
Lower Blue:~Two-mode squeezing in $\hat{x}_1^\text{out}-\hat{x}_2^\text{out}$ and $\hat{p}_1^\text{out}+\hat{p}_2^\text{out}$. 
Upper Blue:~Two-mode antisqueezing in $\hat{x}_1^\text{out}+\hat{x}_2^\text{out}$ and $\hat{p}_1^\text{out}-\hat{p}_2^\text{out}$.
Cyan:~Summed shot noises.
Lower Red:~Theory of two-mode squeezing for optimal PIA outputs.
Upper Red:~Theory of two-mode antisqueezing for optimal PIA outputs.
Magenta:~Theory of two-mode squeezing for our PIA outputs with $-5$~dB squeezed ancillas.
Green:~Theory of two-mode squeezing for our PIA outputs with vacuum ancillas.
}\label{fig:Amp_EPR}
\end{figure}

\begin{figure}[!t]
\centering
\includegraphics[clip,scale=0.478]{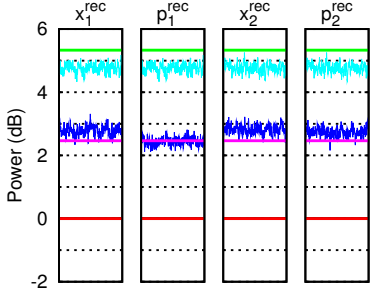}
\caption{
Powers of initial vacuum fluctuations reconstructed from output correlations.
Vertical axes are powers in dB scale normalized by shot noises.
Blue:~Reconstructed quadratures.
Cyan:~Summed shot noises of two homodyne detections.
Red:~Theory for optimal PIA.
Magenta:~Theory for our PIA with $-5$~dB squeezed ancillas.
Green:~Theory for our PIA with vacuum ancillas.
}
\label{fig:Amp_RV}
\end{figure}

\begin{figure}[!t]
\centering
\subfigure[$\avg{\hat{x}_1^\text{in}}\neq0$.]{
\includegraphics[clip,scale=0.478]{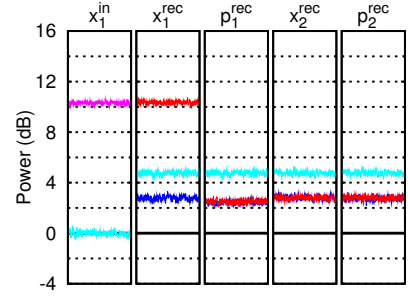}
}
\subfigure[$\avg{\hat{p}_1^\text{in}}\neq0$.]{
\includegraphics[clip,scale=0.478]{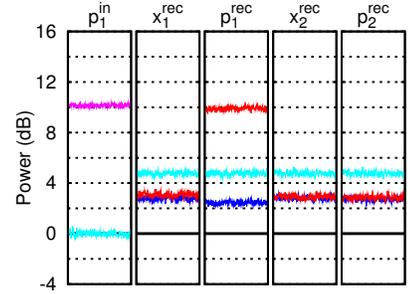}
}
\subfigure[$\avg{\hat{x}_2^\text{in}}\neq0$.]{
\includegraphics[clip,scale=0.478]{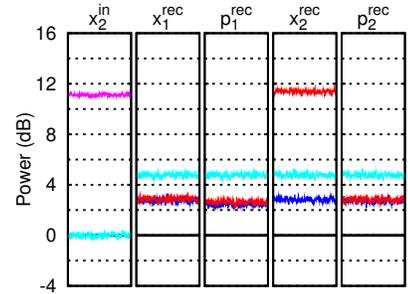}
}
\subfigure[$\avg{\hat{p}_2^\text{in}}\neq0$.]{
\includegraphics[clip,scale=0.478]{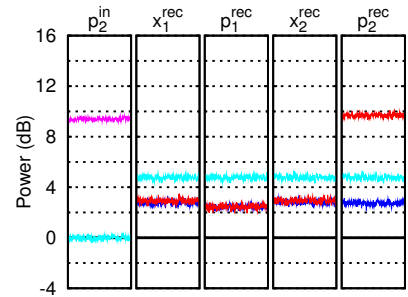}
}
\caption{
Powers of initial quadratures reconstructed from output correlations.
One of four input quadratures $\hat{x}_1^\text{in}$, $\hat{p}_1^\text{in}$, $\hat{x}_2^\text{in}$, $\hat{p}_2^\text{in}$ is excited, at the same time the other three quadratures are left at the vacuum level.
Vertical axes are powers in dB scale normalized by shot noises.
Magenta:~Excited input quadratures.
Red:~Reconstructed quadratures with excitation in inputs.
Blue:~Reconstructed quadratures without excitation in inputs.
Cyan:~Shot noises of a single homodyne detection for the input quadratures and two homodyne detections for the reconstructed quadratures. 
}
\label{fig:Amp_RC}
\vspace{-3\baselineskip}
\end{figure}

In principle, we can fully characterize our PIA with only the above way of analysis that treats quadrature values directly. 
However, such treatment requires a large amount of data for good accuracy.
Thus, in the following, we resort to power measurements using a spectrum analyzer. 
Not only output quadratures (shown in Figs.~\ref{fig:Amp_V} and \ref{fig:Amp_C}) but also their correlations (shown in Figs.~\ref{fig:Amp_EPR}, \ref{fig:Amp_RV}, and \ref{fig:Amp_RC}) are taken for several input states.
In each figure, results of each quadrature are contained in one of boxes.
Vertical axes are powers in dB scale which are normalized by corresponding shot noises. 

Fig.~\ref{fig:Amp_V} shows the experimental results for vacuum inputs (fluctuating traces), together with their theoretical expectations (straight lines).
There are four boxes corresponding to the four output quadratures, namely, $\hat{x}_1^\text{out}$, $\hat{p}_1^\text{out}$, $\hat{x}_2^\text{out}$, and $\hat{p}_2^\text{out}$.
The traces in blue are the powers of the output quadratures. 
The traces in cyan around $0$~dB are the powers of the shot noises, which are used for normalization. 
Since the inputs are in vacuum states, the powers of the shot noises correspond to those of the input quadratures $\hat{x}_1^\text{in}$, $\hat{p}_1^\text{in}$, $\hat{x}_2^\text{in}$, and $\hat{p}_2^\text{in}$.
We put three kinds of theoretical lines corresponding to three different conditions. 
For the optimal PIA of a coherent state with $G=2.0$, the output quadrature variances become three times larger than the initial shot-noise-limited variance, where two from amplification and one from contamination by the other mode. 
The corresponding $4.8$ dB is marked by the lines in red. 
Our PIA with finite ancilla squeezing is suffered from further excess noise.
Assuming $-5$~dB of squeezing for the ancilla states, we calculate theoretical values which are marked by the lines in magenta.
We also show them with vacuum ancillas by the lines in green.
The lines in red, magenta, and green are very close to each other, thus other experimental errors are dominant rather than the ancilla squeezing levels in these results. 
Therefore, it is hard to discuss nonclassicality of our PIA only from these results. 
As is shown later, the effects of ancilla squeezing more clearly appear in the output correlations. 

Next we use several coherent states as inputs. 
The results are shown in Fig.~\ref{fig:Amp_C}. 
The four input quadratures $\hat{x}_1^\text{in}$, $\hat{p}_1^\text{in}$, $\hat{x}_2^\text{in}$, and $\hat{p}_2^\text{in}$ are displaced from zero mean values by turns, leaving the other three quadratures at the vacuum level. 
There are four subfigures labeled from~(a) to~(d) corresponding to such four excitations. 
For each subfigure, there are five boxes. 
The trace in magenta in the leftmost box shows the measured power of the excited input quadrature. 
The other four boxes correspond to the four output quadratures, namely, $\hat{x}_1^\text{out}$, $\hat{p}_1^\text{out}$, $\hat{x}_2^\text{out}$, and $\hat{p}_2^\text{out}$.
The traces in red show the output quadrature powers with the input excitation.
They are compared to those without the excitation shown by the blue traces, which are replottings of the blue traces in Fig.~\ref{fig:Amp_V}. 
The obtained results show the following features. 
When a quadrature $\hat{x}$ or $\hat{p}$ of an input mode is excited, the same quadratures of both of the output modes are excited, whereas the conjugate quadratures do not change from the nonexcited levels. 
The two increased output powers differ by about $3.0$~dB, where the larger one corresponds to the amplified signal and the smaller one corresponds to the phase-conjugated idler output. 
These features are exactly what are expected from Eq.~\eqref{eq:PiaUnitary} for $G=2$ and $\theta=0$. 
Note that the coefficients $\sqrt{2}$ correspond to the $3.0$~dB. 

The results in Figs.~\ref{fig:Amp_V} and \ref{fig:Amp_C} are only for the five specific input states. 
However, the results for other input states can be predicted on the assumption of linearity. 
More precisely speaking, the absolute values of the coefficients of $\hat{x}_1^\text{in}$, $\hat{p}_1^\text{in}$, $\hat{x}_2^\text{in}$, and $\hat{p}_2^\text{in}$ in Eq.~\eqref{eq:PiaUnitary} are determined from these results. 
On the other hand, the signs of the coefficients are not determined from them. 
However, they are checked from the phase space diagrams shown in Figs.~\ref{fig:Amp_PS1} and \ref{fig:Amp_PS2}. 
In the above sense, the results shown so far give full information of the input-output relation when output modes are separately concerned. 

In order to fully characterize our amplifier, the individual behaviors of the output modes are not enough. 
In the following, we are concerned with the output correlations. 

Since unitary PIA is equivalent to two-mode squeezing, the two output modes should be entangled, and have an EPR type of correlation. 
The results of EPR correlation are shown in Fig.~\ref{fig:Amp_EPR}. 
Here the two input modes are both in vacuum states. 
There are two boxes corresponding to $x$ and $p$ correlations. 
The lower traces in blue show the two-mode squeezing of $\hat{x}_1^\text{out}-\hat{x}_2^\text{out}$ and $\hat{p}_1^\text{out}+\hat{p}_2^\text{out}$, on the other hand, the upper traces in blue show the two-mode antisqueezing of $\hat{x}_1^\text{out}+\hat{x}_2^\text{out}$ and $\hat{p}_1^\text{out}-\hat{p}_2^\text{out}$, respectively. 
They are compared with the summed shot noises of the two homodyne detections shown by the traces in cyan. 
Several theoretical lines are plotted together. 
The lower and upper lines in red are the theoretical values of two-mode squeezing and antisqueezing for the optimal PIA, respectively.
Our results of two-mode squeezing are degraded from the ideal case due to finite squeezing of ancillas. 
Assuming $-5$~dB squeezing for ancillas, theoretical expectation is marked by the lines in magenta. 
That for vacuum ancillas is also marked by the lines in green, which is exactly equal to the shot noise level. 
In contrast, the theoretical two-mode antisqueezing is always ideal for arbitrary ancillas. 
The experimental results well agree with the theory assuming $-5$~dB squeezing of ancillas.
Since the lower traces in blue are both below the traces in cyan, existence of entanglement is verified between the two output modes via the Duan-Simon criterion~\cite{Duan(2000):PRL,Simon(2000):PRL}.

From the nonclassical correlation between the two output modes, we investigate the reversibility of our PIA. 
For this purpose, we virtually realize the inverse transformation electrically and reconstruct the initial quadratures. 
Neglecting the excess noise from finite ancillas, PIA that we demonstrate has the input-output relation as $\hat{a}_1^\text{out}=\sqrt{2}\hat{a}_1^\text{in}+(\hat{a}_2^\text{in})^\dagger$, $\hat{a}_2^\text{out}=\sqrt{2}\hat{a}_2^\text{in}+(\hat{a}_1^\text{in})^\dagger$, which is obtained by substituting $G=2$ and $\theta=0$ into Eq.~\eqref{eq:PiaUnitary}.
The inverse transformation becomes as
$\hat{a}_1^\text{out}=\sqrt{2}\hat{a}_1^\text{in}-(\hat{a}_2^\text{in})^\dagger$, 
$\hat{a}_2^\text{out}=\sqrt{2}\hat{a}_2^\text{in}-(\hat{a}_1^\text{in})^\dagger$, 
or equivalently, 
\begin{subequations}\label{eq:PiaInv}
\begin{align}
\hat{x}_1^\text{out}= & \sqrt{2}\hat{x}_1^\text{in}-\hat{x}_2^\text{in}, &
\hat{x}_2^\text{out}= & \sqrt{2}\hat{x}_2^\text{in}-\hat{x}_1^\text{in}, \\
\hat{p}_1^\text{out}= & \sqrt{2}\hat{p}_1^\text{in}+\hat{p}_2^\text{in}, &
\hat{p}_2^\text{out}= & \sqrt{2}\hat{p}_2^\text{in}+\hat{p}_1^\text{in}. 
\end{align}
\end{subequations}
Therefore, by adding or subtracting the two homodyne outcomes with $3.0$~dB difference of gains, the initial quadratures are reconstructed. 
The reconstructed quadratures are denoted by the superscripts ``rec'' in the following. 
Note that the initial quantum state is not recovered in the experiment. 
In addition, note also that only one of two quadratures $\hat{x}$ or $\hat{p}$ can be reconstructed in each moment, and never both simultaneously. 
The recovery of the initial state is possible only when either a quantum channel is between the signal and idler outputs~\cite{Filip(2004):PRA} or linear pre-processing is applied~\cite{Radim(2009):PRA}. 
However, the demonstrated reconstruction of the initial quadratures can show that necessary correlations for the recovery of the initial state are present.

In Fig.~\ref{fig:Amp_RV}, the results of such reconstruction of initial vacuum fluctuations are shown. There are four boxes corresponding to the four reconstructed quadratures $\hat{x}_1^\text{rec}$, $\hat{p}_1^\text{rec}$, $\hat{x}_2^\text{rec}$, and $\hat{p}_2^\text{rec}$.
The powers of the reconstructed quadratures are shown by the traces in blue.
The traces in cyan are the powers of the summed shot noises of the two homodyne detections, which are taken with the same electric gains as the traces in blue.
Note that the lower levels of the blue traces than the cyan traces are due to the nonclassical correlation. 
From Eq.~\eqref{eq:PiaInv}, the summed shot noises should have three times larger variances than those corresponding to the initial vacuum fluctuations. 
Thus, we infer the original vacuum level to be $4.8$~dB below the measured sum of the shot noises. 
All results shown here are normalized by the inferred vacuum level. 
For the optimal PIA, vacuum fluctuations are perfect reconstructed, thus the theoretical expectation coincides with the vacuum level of $0$~dB, which is marked by the lines in red. 
The increases of the blue traces from $0$~dB show the imperfection of our PIA. 
The theoretical values for $-5$~dB squeezing of and no squeezing of ancillas are marked by the lines in magenta and green, respectively.
The experimental results of the traces in blue are in good agreement with the lines in magenta. 

Next, we pay attention to the reconstruction of mean amplitude, using coherent states as inputs. 
The results are shown in Fig.~\ref{fig:Amp_RC}.
The four input quadratures, namely $\hat{x}_1^\text{in}$, $\hat{p}_1^\text{in}$, $\hat{x}_2^\text{in}$, and $\hat{p}_2^\text{in}$, are excited one by one, as shown in the leftmost boxes of four subfigures. 
For each excitation, the powers of the four reconstructed quadratures are measured, the results of which are put in the other four boxes on the right side. 
For the leftmost box, the trace in magenta is the power of the excited input quadrature, and the trace in cyan is the power of the shot noise of the corresponding homodyne detection. 
The increase of the magenta trace from the cyan trace indicates the excitation. 
For the other four boxes, the traces in red and blue show the powers of the reconstructed quadratures with and without the input excitation respectively, and the traces in cyan are the summed shot noise powers of the two homodyne detections. 
Similarly to Fig.~\ref{fig:Amp_RV}, one third of the summed shot noise power is used for normalization. 
The reconstructed quadratures are excited almost to the same levels as the input quadratures at the excited quadratures, whereas they remain unchanged from the nonexcited levels at the nonexcited quadratures. 

All the results shown above prove the success of our demonstration of PIA. 
They agree well with the theoretical calculations assuming $-5$~dB of squeezing for ancillas.

\section{Experimental Results for Cloner}
\label{sec:ResultsClone}

We show next the results of the cloning experiment in a manner similar to the PIA experiment, i.e., quadrature data and the phase space diagrams reconstructed from them are used for intuitive understanding and also for check of the mirror image that is found in the anticlone, and the full verification is given by the power analysis for various input states. 
In the following, since the representation and interpretation of results are almost the same as those for the PIA experiment, we give only short description of them. 
The signal input of PIA is denoted by the term ``original'', and the resulting two clones and one anticlone are denoted by ``clone-1'', ``clone-2'', and ``anticlone'', respectively.
There are also abbreviation of these terms as ``org'', ``cln-1'', ``cln-2'', and ``a-cln'' in the figures and mathematical expressions. 
Occasionally, the term ``input'' is used indicating the original, and ``output'' indicating the two clones and one anticlone.

\begin{figure}[!tb]
\centering
\subfigure[Original.]{
\includegraphics[clip,scale=0.319]{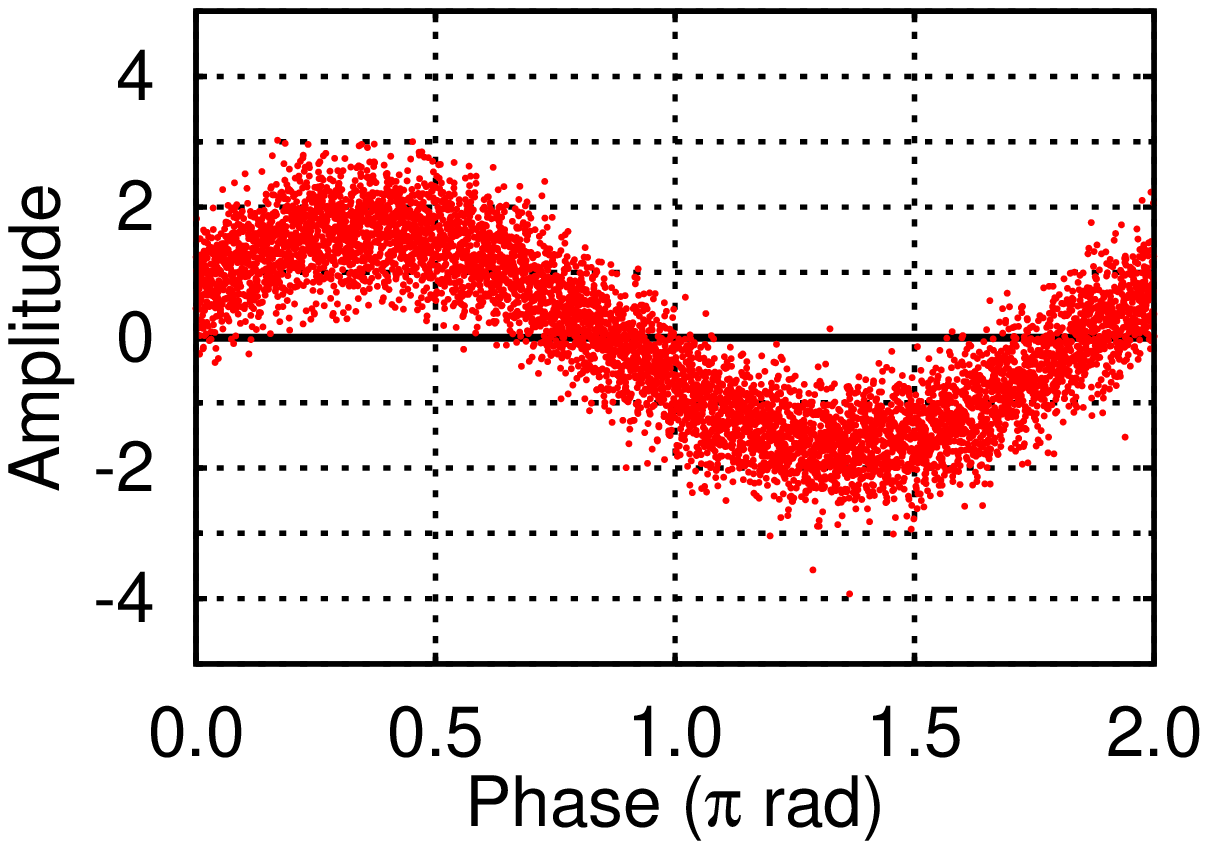}
\label{sfig:Cln_In}
}
\subfigure[Clone-1.]{
\includegraphics[clip,scale=0.319]{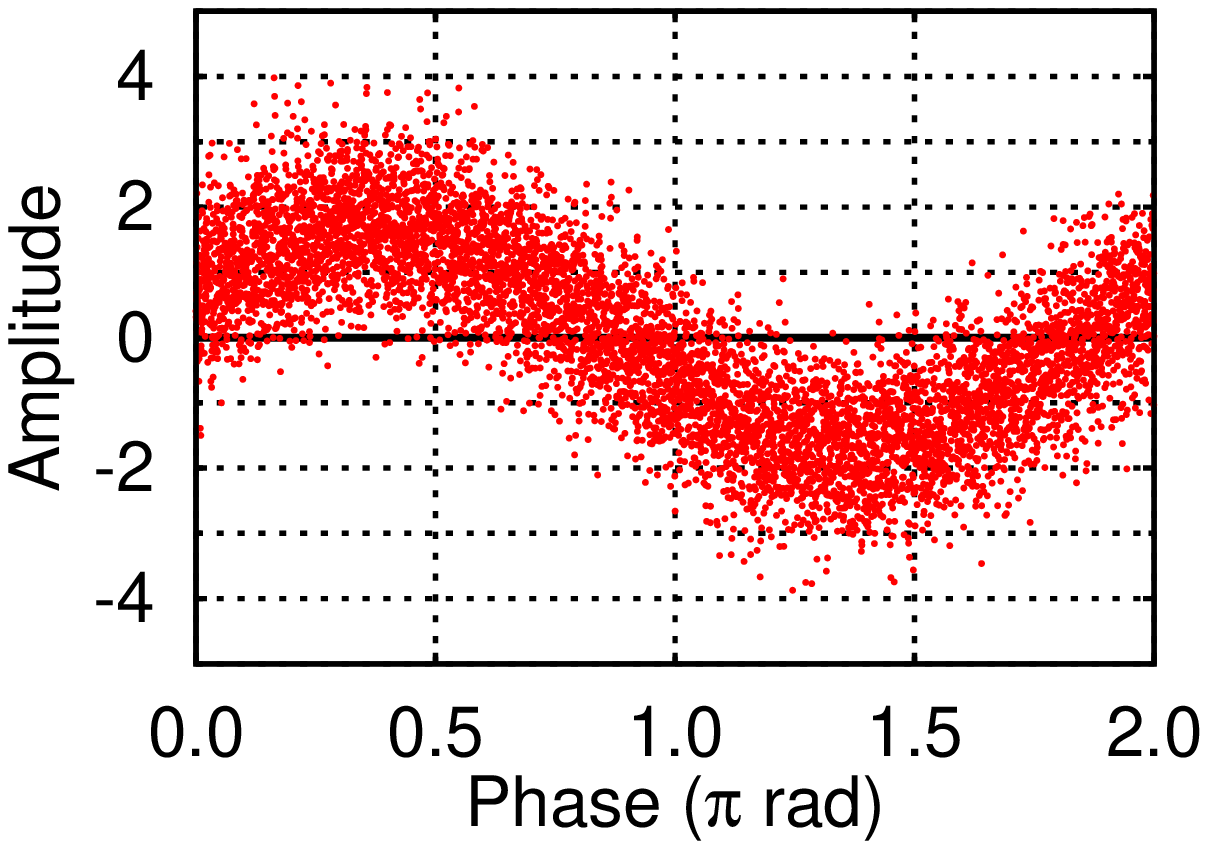}
\label{sfig:Cln_Cln1}
}\\
\subfigure[Clone-2.]{
\includegraphics[clip,scale=0.319]{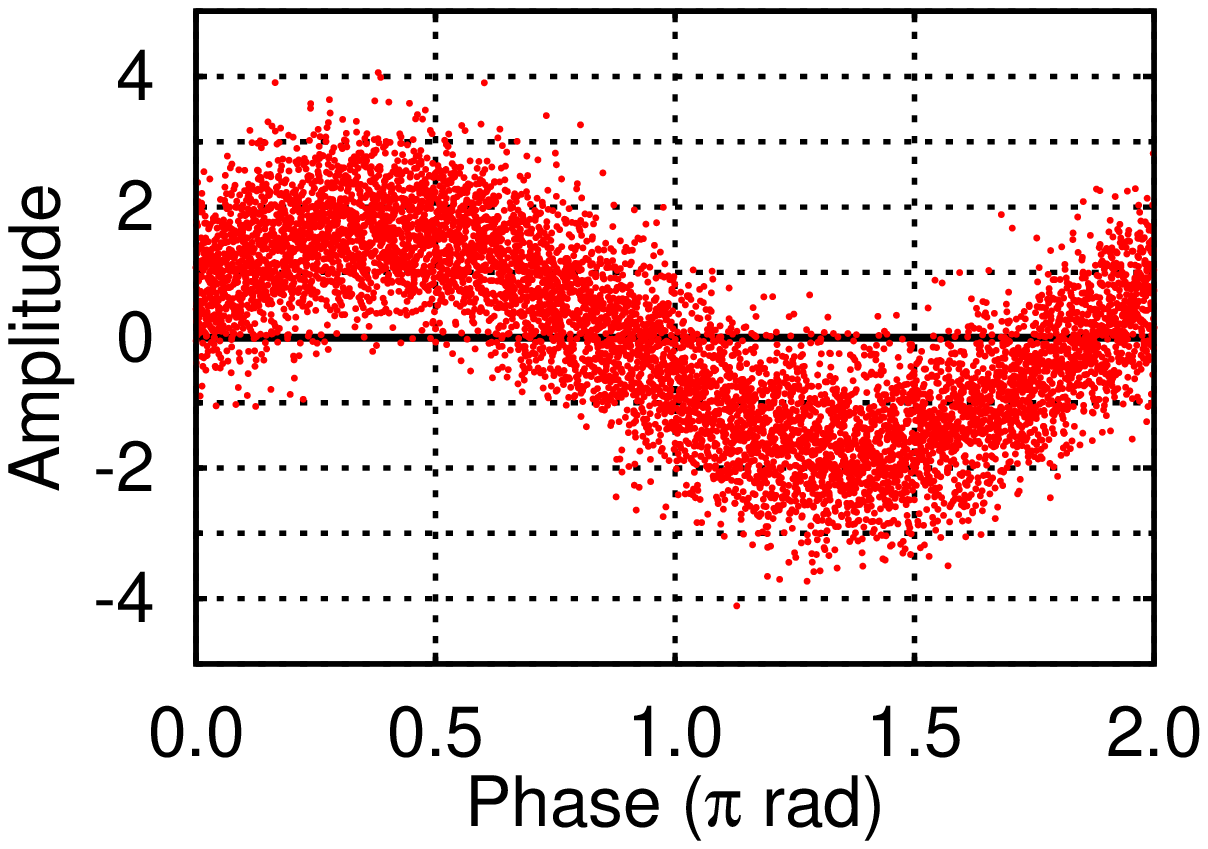}
\label{sfig:Cln_Cln2}
}
\subfigure[Anticlone.]{
\includegraphics[clip,scale=0.319]{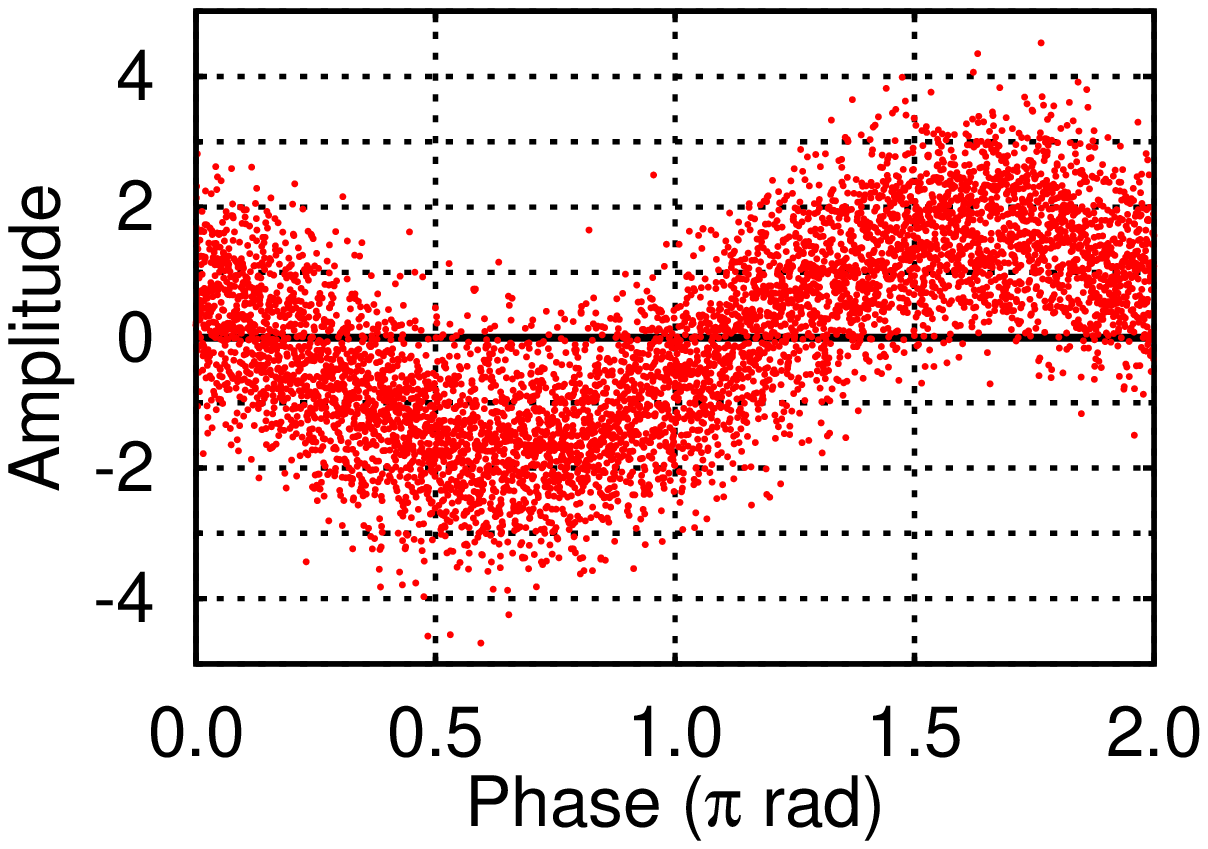}
\label{sfig:Cln_AntiCln}
}
\caption{
Quadrature data. 
The phases of homodyne measurements are scanned from $0$ to $2\pi$, which correspond to horizontal axes.
Vertical axes are quadrature values.
}\label{fig:Cln_MD}
\subfigure[Experiment.]{
\includegraphics[clip,scale=0.45]{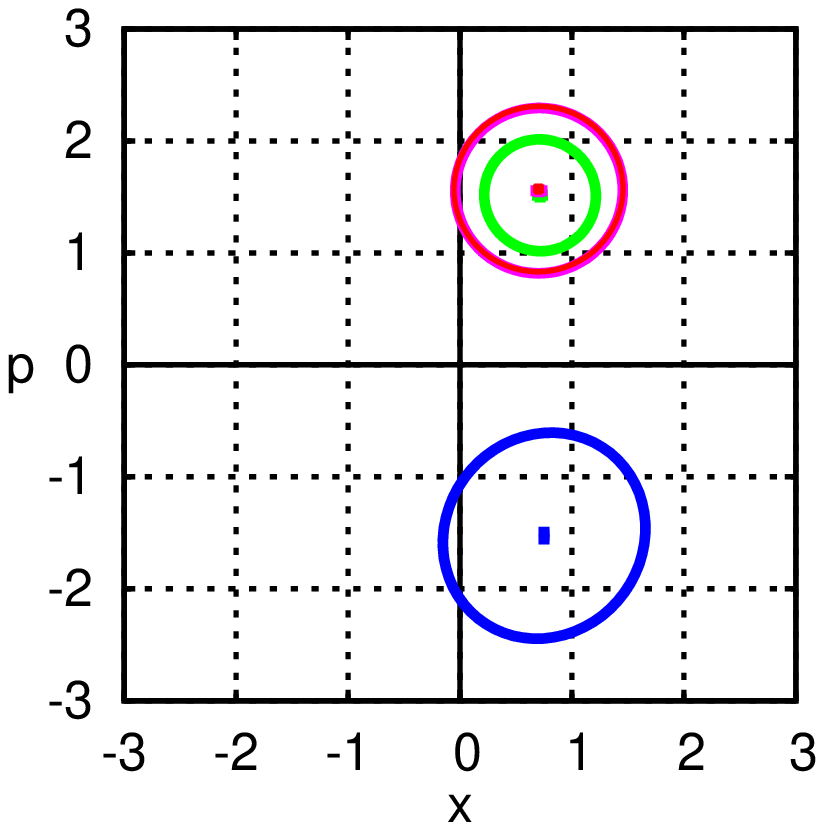}
}
\subfigure[Theory.]{
\includegraphics[clip,scale=0.45]{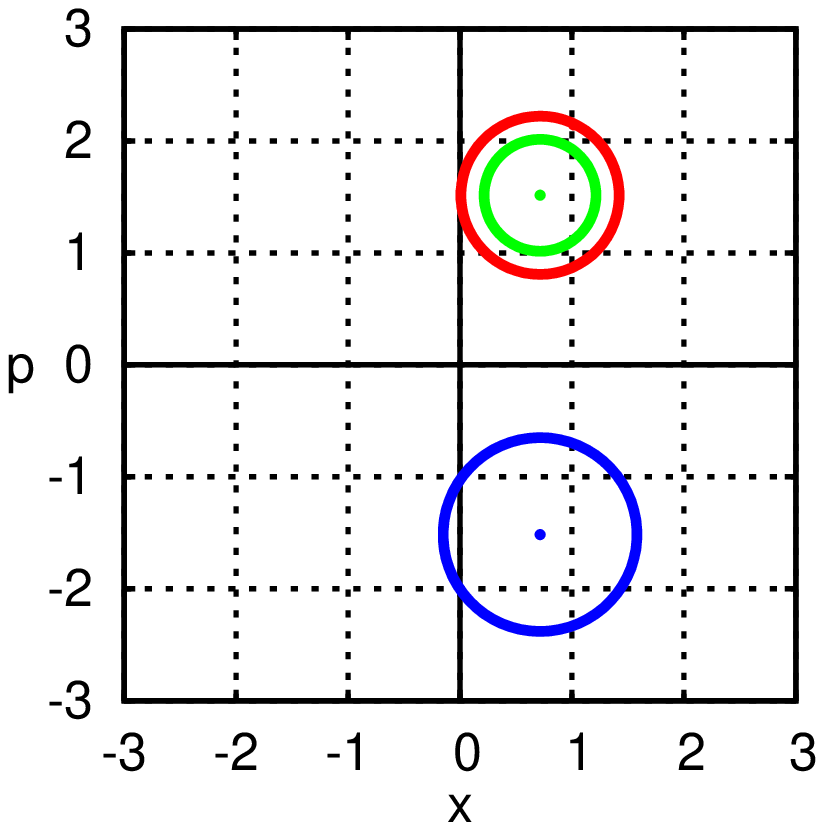}
}
\caption{
Phase space distributions computed from the quadrature data in Fig.~\ref{fig:Cln_MD}.
The first and second moments of Gaussian Wigner functions are expressed by ellipses.
Green:~Original.
Red:~Clone-1.
Magenta:~Clone-2.
Blue:~Anticlone.
}\label{fig:Cln_PS}
\end{figure}

Fig.~\ref{fig:Cln_MD} shows the quadrature data. 
From an original~(a) in a coherent state, two clones~(b) and (c) and an anticlone~(d) are produced. 
We see that the original and the two clones have almost the same sinusoidal curves of the mean amplitudes, though the fluctuations are uniformly increased in the clones. 
On the other hand, the anticlone has the same sinusoid when the phase is flipped. 

Fig.~\ref{fig:Cln_PS} is the phase space diagram computed from the quadrature data shown in Fig.~\ref{fig:Cln_MD}. 
The first and second moments of each distribution are represented by an ellipse.
Next to the experimental diagram~(a), theoretical calculations for the optimal cloning is depicted~(b). 
The two ellipses of the clones (red and magenta) in the experimental diagram are almost overlapped. 
The centers of the two clones are almost the same as that of the original~(green), whereas the radii of the clones are larger than that of the original. 
On the other hand, the anticlone~(blue) has a different center, where the sign of $p$ quadrature is opposite to that of the original. 

\begin{figure}[!tb]
\centering
\includegraphics[clip,scale=0.478]{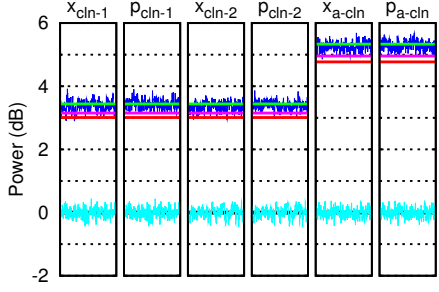}
\caption{
Output powers for vacuum originals.
Vertical axes are powers in dB scale normalized by shot noises.
Blue:~Output quadratures.
Cyan:~Shot noises.
Red:~Theory for optimal cloner.
Magenta:~Theory for our cloner with $-5$~dB squeezed ancillas.
Green:~Theory for our cloner with vacuum ancillas.
}\label{fig:Cln_OV}
\end{figure}

\begin{figure}[!tb]
\centering
\subfigure[$\avg{\hat{x}_\text{org}}\neq0$.]{
\includegraphics[clip,scale=0.478]{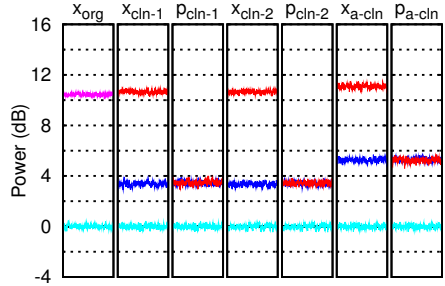}
}
\subfigure[$\avg{\hat{p}_\text{org}}\neq0$.]{
\includegraphics[clip,scale=0.478]{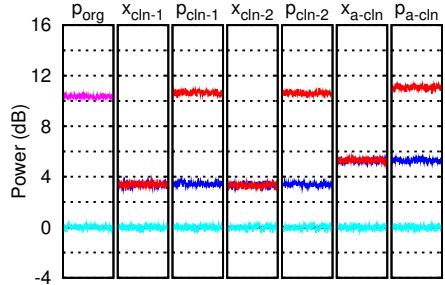}
}
\caption{
Output powers for originals in several coherent states.
One of two original quadratures $\hat{x}_\text{org}$, $\hat{p}_\text{org}$ is excited, at the same time the other quadrature is left at the vacuum level.
Vertical axes are powers in dB scale normalized by shot noises.
Magenta:~Excited original quadratures.
Red:~Output quadratures with excitation in originals.
Blue:~Output quadratures without excitation in originals.
Cyan:~Shot noises.
}\label{fig:Cln_OC}
\end{figure}

\begin{figure}[!tb]
\centering
\subfigure[EPR correlation between clone-1 and anticlone.]{
\hspace{0.05\textwidth}
\includegraphics[clip,scale=0.478]{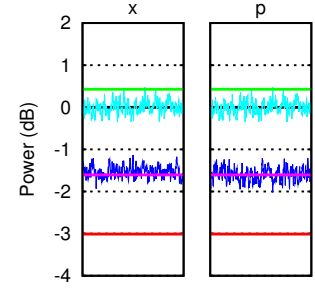}
\label{sfig:Cln_EPR_1}
\hspace{0.05\textwidth}
}
\subfigure[EPR correlation between clone-2 and anticlone.]{
\hspace{0.05\textwidth}
\includegraphics[clip,scale=0.478]{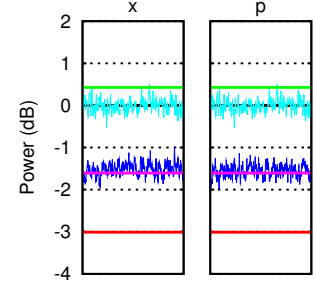}
\label{sfig:Cln_EPR_2}
\hspace{0.05\textwidth}
}
\caption{
EPR correlation between each clone and the anticlone.
Vertical axes are powers in dB scale normalized by summed shot noises of two homodyne detections.
Blue:~$\hat{x}_\text{cln}-\hat{x}_\text{a-cln}$ and $\hat{p}_\text{cln}+\hat{p}_\text{a-cln}$.
Here the subscripts ``cln'' denote clone-1 for (a) and clone-2 for (b).
Cyan:~Summed shot noises of two homodyne detections.
Red:~Theory for optimal cloner.
Magenta:~Theory for our cloner with $-5$~dB squeezed ancillas.
Green:~Theory for our cloner with vacuum ancillas.
}\label{fig:Cln_EPR}
\end{figure}

\begin{figure}[!tb]
\centering
\includegraphics[clip,scale=0.478]{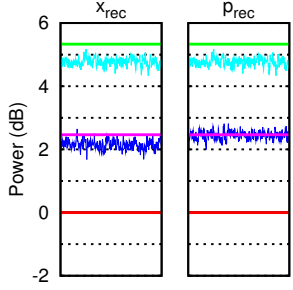}
\caption{
Powers of initial vacuum fluctuations reconstructed from output correlations. 
Vertical axes are powers in dB scale normalized by shot noises. 
Blue:~Reconstructed quadratures.
Cyan:~Summed shot noises of three homodyne detections.
Red:~Theory for optimal cloner.
Magenta:~Theory for our cloner with $-5$~dB squeezed ancillas.
Green:~Theory for our cloner with vacuum ancillas.
}\label{fig:Cln_RV}
\end{figure}

\begin{figure}[!tb]
\centering
\subfigure[$\avg{\hat{x}_\text{org}}\neq0$.]{
\includegraphics[clip,scale=0.478]{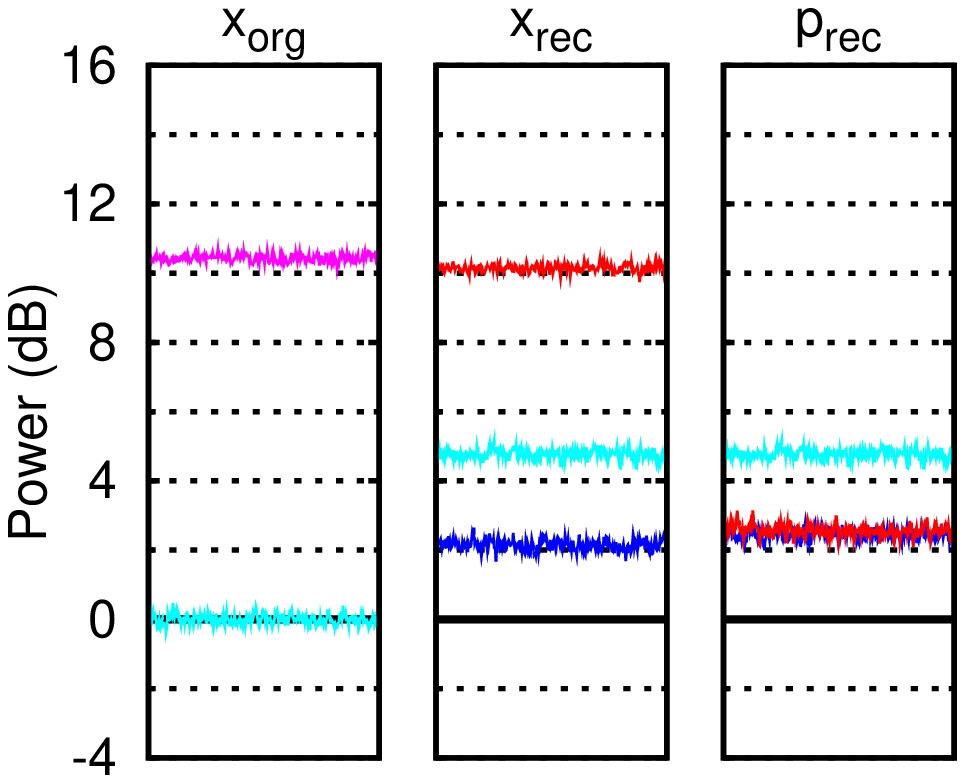}
}
\subfigure[$\avg{\hat{p}_\text{org}}\neq0$.]{
\includegraphics[clip,scale=0.478]{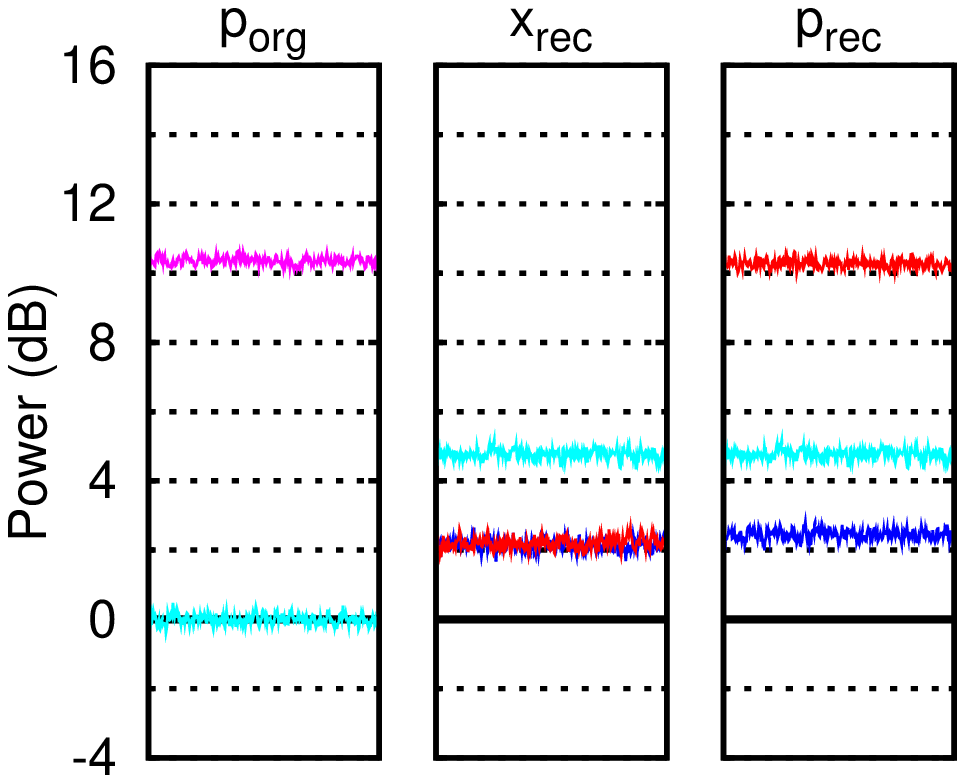}
}
\caption{
Powers of initial quadratures reconstructed from output correlations.
One of two original quadratures $\hat{x}_\text{org}$, $\hat{p}_\text{org}$ is excited, at the same time the other quadrature is left at the vacuum level.
Vertical axes are powers in dB scale normalized by shot noises.
Magenta:~Excited original quadratures.
Red:~Reconstructed quadratures with excitation in originals.
Blue:~Reconstructed quadratures without excitation in originals.
Cyan:~Shot noises of a single homodyne detection for the input quadratures and three homodyne detections for the reconstructed quadratures.
}\label{fig:Cln_RC}
\end{figure}

We move on to the power analysis. 
First we show the powers of the output quadratures (shown in Figs.~\ref{fig:Cln_OV} and \ref{fig:Cln_OC}), and then their correlations (shown in Figs.~\ref{fig:Cln_EPR}, \ref{fig:Cln_RV}, and \ref{fig:Cln_RC}).
Fig.~\ref{fig:Cln_OV} shows the cloning of a vacuum state, and Fig.~\ref{fig:Cln_OC} shows the cloning of several coherent states. 

In Fig.~\ref{fig:Cln_OV}, there are six boxes corresponding to six output quadratures, namely, $\hat{x}_\text{cln-1}$, $\hat{p}_\text{cln-1}$, $\hat{x}_\text{cln-2}$, $\hat{p}_\text{cln-2}$, $\hat{x}_\text{a-cln}$, and $\hat{p}_\text{a-cln}$.
For each box, there are two experimental traces.
The traces in blue are the powers of the output quadratures. 
The traces in cyan are the shot noise powers used for normalization, which are equal to the powers of the input quadratures.
There are also three kinds of theoretical lines. 
The lines in red are for the optimal cloning, and the lines in magenta and green are for our cloning using $-5$~dB squeezed and vacuum ancillas for PIA, respectively. 
Note that the lines in red at $3.0$~dB for clones correspond to the cloning limit. 
From these results, cloning fidelities are estimated at $F=0.63\pm0.01$ for a vacuum original, which is higher enough than the classical limit of $F=1/2$ and very close to the cloning limit of $F=2/3$. 

In Fig.~\ref{fig:Cln_OC}, there are two subfigures (a) and (b) corresponding to excitations in $\hat{x}_\text{org}$ and $\hat{p}_\text{org}$, respectively.
Each subfigure is composed of seven boxes, where the leftmost one shows the input excitation in $\hat{x}_\text{org}$ or $\hat{p}_\text{org}$, and the other six boxes show the output quadratures of $\hat{x}_\text{cln-1}$, $\hat{p}_\text{cln-1}$, $\hat{x}_\text{cln-2}$, $\hat{p}_\text{cln-2}$, $\hat{x}_\text{a-cln}$, and $\hat{p}_\text{a-cln}$.
The trace in magenta shows the power of the excited input quadrature. 
The traces in red and blue show the output powers with and without the input excitation, respectively. 
The traces in cyan are the shot noise powers used for normalization.
When we excite one quadrature of the original, the same quadratures in the three output modes are excited to almost the same level, whereas the conjugate quadratures do not change.

Our remaining concerns are the output correlations and the reversibility.
First, in Fig.~\ref{fig:Cln_EPR}, we show EPR correlation between each clone and the anticlone as a sufficient condition for entanglement. 
The correlation between the clone-1 and the anticlone is shown in Fig.~\ref{sfig:Cln_EPR_1}, and that between the clone-2 and the anticlone is shown in Fig.~\ref{sfig:Cln_EPR_2}. 
By electrically adding or subtracting the two homodyne signals with the same electric gains, four observables are measured, namely, $\hat{x}_\text{cln-1}-\hat{x}_\text{a-cln}$, $\hat{p}_\text{cln-1}+\hat{p}_\text{a-cln}$, $\hat{x}_\text{cln-2}-\hat{x}_\text{a-cln}$, and $\hat{p}_\text{cln-2}+\hat{p}_\text{a-cln}$, which are separately contained in boxes and shown as the blue traces. 
These traces are all below the summed shot noises of the traces in cyan. 
From these results, we verify bipartite entanglement between each clone and the anticlone from Duan-Simon criterion~\cite{Duan(2000):PRL,Simon(2000):PRL}, and eventually, tripartite entanglement of Class I where none of three partial systems is separable from the others~\cite{Giedke(2001):PRA}. 
Theoretical lines in red, magenta, and green are plotted together, corresponding to infinite squeezing, finite squeezing of $-5$~dB, and no squeezing of ancillas for PIA, respectively. 
The experimental results of the traces in blue well agree with the lines in magenta.

Using the nonclassical correlations, we reconstruct the original quadratures. 
The reconstructed quadratures are denoted by $\hat{x}_\text{rec}$ and $\hat{p}_\text{rec}$.
The results for a vacuum state are shown in Fig.~\ref{fig:Cln_RV}, and those for coherent states are shown in Fig.~\ref{fig:Cln_RC}. 
For the reconstruction, three homodyne signals are added with the same electric gains and appropriate signs. 
For the similar reason to the PIA experiment, the summed shot noise has three times larger variance than that corresponding to the vacuum fluctuation of the original. 
Thus, one third of the summed shot noise power is used for normalization.

In Fig.~\ref{fig:Cln_RV}, the powers of the reconstructed vacuum fluctuations are plotted as the traces in blue, and compared to that of the summed shot noise plotted as the traces in cyan. 
The blue traces are below the cyan traces for both $\hat{x}_\text{rec}$ and $\hat{p}_\text{rec}$ due to nonclassical correlations. 
Theoretical expectations are also shown as the lines in red, magenta, and green corresponding to the three different conditions of infinite squeezing, finite squeezing of $-5$~dB, and no squeezing of ancillas, respectively. 
The perfect reconstruction corresponding to $0$~dB is marked by the red lines, which is not achieved in the experiment due to the finite squeezing of ancillas. 
The results are degraded almost to the level of the magenta lines, as expected from the theory. 
However, they are still slightly below $3.0$~dB which corresponds to the cloning limit. 
From these results, the fidelity of reconstruction is calculated for a vacuum state. 
A perfect unitary cloning allows the reconstruction fidelity of $F=1$. 
The experimental value is calculated as $F=0.74\pm0.01$, which is higher than the cloning limit of $F=2/3$. 
The cloning limit can be considered as the classical limit for the reproduction of the original state, because one can never obtain a better approximation of the original state than the clones if the nonclassical correlations between the clones and anticlones can not be utilized. 

In Fig.~\ref{fig:Cln_RC}, the input quadratures $\hat{x}_\text{org}$ and $\hat{p}_\text{org}$ are excited one by one, and for each case they are reconstructed from the output correlations. 
The trace in magenta in the leftmost box is the power of the excited original quadrature, whereas the trace in cyan in the same box is the power of the corresponding shot noise. 
The traces in red and blue in the other two boxes on the right side are the powers of the reconstructed quadratures with and without the excitation, respectively, and the traces in cyan at $4.8$~dB are the powers of the summed shot noises of the three homodyne detections. 
At the excited quadrature, almost the same level of excitation is reconstructed. 
In contrast, at the conjugate quadrature, no effect of the excitation is observed.

\section{Summary}
\label{sec:Summary}

We succeeded in phase-insensitive optical amplification in a reversible manner. 
Our amplifier preserves the idler output. 
The entanglement between the signal and idler is responsible for the reversibility. 
The scheme is basically based on linear optics, homodyne measurements and feedforward. 
Offline-prepared squeezed states which are used as ancillas provide nonclassical properties for our PIA. 
We demonstrated for the amplification gain of $G=2.0$. 
By splitting the amplified output in half, we also demonstrated $1\to2$ approximate cloning of coherent states, where the remaining idler output was interpreted as the anticlone. 

For both experiments, the full demonstration was given in the following sequence.
First, we characterized the individual output modes. 
By treating the quadrature data directly, they were visualized as phase space diagrams to help intuitive understanding. 
Especially, the mirror image in the idler output or the anticlone is shown. 
Then the input-output relation was examined more strictly by using several different coherent states as inputs. 
Finally, the output correlations were examined. 
They are important because the nonclassical properties are only accessible via them.
Not only the ordinary EPR correlations were shown, but also the possibility of the reverse operation was directly presented by appropriate measurements of the correlations. 

Our results are a good demonstration that shows the properties of an amplification process, which have been theoretically known for decades but not fully demonstrated experimentally. 
Especially, it is reversible when the idler is present. 
Such reversible amplification is significant from practical respects, as is shown in Refs.~\cite{Radim(2009):PRA,Filip(2004):PRA}. 
We did not demonstrate the reverse operation but showed its possibility from the correlations. 
The recovery of the signal state only requires the Bell measurement and feedforward~\cite{Radim(2009):PRA,Filip(2004):PRA}, which would be much less lossy and noisy than implementing the inverse transformation. 
Full and partial recovery of the distributed information via such a feedforward scheme is left for future experiments.

\section*{Acknowledgement}

This work was partly supported by SCF, GIA, G-COE, PFN and FIRST commissioned by the MEXT of Japan, the Research Foundation for Opt-Science and Technology, SCOPE program of the MIC of Japan, and, 
JSPS and ASCR under the Japan-Czech Republic Research Cooperative Program.
R.~F. acknowledges projects: MSM 6198959213 and ME10156 of the Czech Ministry of Education, 
grant 202/08/0224 of GA \v CR and EU Grant FP7 212008 COMPAS.

\appendix
\section{General number of clones}
\label{sec:GenCln}

In this appendix, the discussion in Sec.~\ref{sec:Clone} is extended to ${K}\to{L}$ cloning. 

The procedure of ${K}\to{L}$ symmetric cloning can be decomposed into three steps as follows~\cite{Braunstein(2001):PRL,Fiurasek(2001):PRL}. 
Firstly, all the information of $(x_\text{d},p_\text{d})$ are put together into a single mode by a beamsplitter network. 
A state with larger amplitude $\hat{D}(\sqrt{K}x_\text{d},\sqrt{K}p_\text{d})\ket{\psi}$ is created from the $K$ identical originals $\hat{D}(x_\text{d},p_\text{d})\ket{\psi}$ in this step. 
Secondly, the combined signal is amplified with the gain $G=L/K$. 
Finally, the amplified signal is combined with $L-1$ ancillas by another beamsplitter network, creating $L$ clones. 
For asymmetric cloning, the procedure is essentially the same, but the amplification in the second step is applied to a part of the combined signal, and the gain is changed correspondingly as $G=1+\sum_{k=1}^Ln_k$~\cite{Fiurasek(2007):PRA}. 
On the other hand, $L-K$ anticlones are obtained from the idler output by combining it with $L-K-1$ ancillas by another beamsplitter network. 
Therefore, as a whole, $L$ clones and $L-K$ anticlones are obtained, whose complex mean amplitudes are $\alpha$ and $\alpha^\ast$ respectively, from $K$ originals with the complex mean amplitude of $\alpha$. 
It is evident that there is no entanglement among clones or anticlones, however, there is entanglement between a clone and an anticlone. 

The asymmetric clones have the following form: 
\begin{align}
\hat{a}_{\text{cln-}k}=\tfrac{1}{\sqrt{K}}\hat{a}_\text{org}^\prime+\sqrt{n_k}\hatd{a}_\text{idl}+\sum_{k=1}^{L-1}\kappa_{k\ell}\hat{a}_{\text{anc-}\ell},
\label{eq:GenClnAsym}
\end{align}
where $\hat{a}_{\text{cln-}k}$ is the annihilation operator of the \mbox{$k$-th} clone, $\hat{a}_\text{org}^\prime$ is that of the combined original after the first step, $\hat{a}_\text{idl}$ is that of the idler input, and $\hat{a}_{\text{anc-}\ell}$ is that of the \mbox{$\ell$-th} ancilla input for the latter beamsplitter network. 
The coefficients $\sqrt{n_k}$ and $\kappa_{k\ell}$ are not independent in order to preserve the commutation relations. 
By setting all the ancillas in vacuum states, the added noises become rotationally symmetric and Gaussian, and their variances correspond to the parameters $n_k$~\cite{Note}. 
They satisfy the relation below~\cite{Fiurasek(2007):PRA}: 
\begin{align}
\Bigl(\sum_{k=1}^L\sqrt{n_k}\Bigr)^2=(L-K)\Bigl(\sum_{k=1}^L n_k+1\Bigr). 
\label{eq:GenClnAsymNoise}
\end{align}
The optimality of Eq.~\eqref{eq:GenClnAsymNoise} is proven with respect to the cost function constructed from the variances~\cite{Fiurasek(2007):PRA}: 
\begin{gather}
C(n_1,\dots,n_L)=\sum_{k=1}^Lc_kn_k. 
\end{gather}
In particular, for the symmetric cloning, i.e., $n_1=\dots=n_L\equiv{n}$, Eq.~\eqref{eq:GenClnAsymNoise} saturates the following inequality which is obtained in Ref.~\cite{Cerf(2000):PRA} from the consistency with the uncertainty relation: 
\begin{align}
n \ge \bigl(\tfrac{1}{K}-\tfrac{1}{L}\bigr). 
\end{align}

The limit fidelity for symmetric Gaussian cloning is calculated as $F=KL/(KL-K+L)$. 
On the other hand, taking the limit that $L$ goes to infinity, the classical limit of cloning (i.e., the limit of state estimation) is obtained as $F=K/(K+1)$. 

As with $1\to2$ cloning, optimal $1\to{L}$ cloning can be fully reversed by $L-1$ Bell measurements performed on each set of a clone and an anticlone and subsequent feedforward to the remaining single clone~\cite{Filip(2004):PRA}. 
Even with the smaller number of Bell measurements, the original is partially recovered accordingly.


\begin{thebibliography}{99}
\bibitem{Note}
More strictly speaking, the story becomes as follows: 
The simple picture of cloning that some noise is added to the original $\hat{D}(x_\text{d},p_\text{d})\ket{\psi}$ is valid for $1\to{L}$ cloning, irrespective of the core state $\ket{\psi}$. 
However, for ${K}\to{L}$ cloning with $K\ge2$, it is not always valid. 
When the core state $\ket{\psi}$ is known, there would be some choice of the ancilla states which make the picture valid. 
Indeed, when the core state is known to be a vacuum state (i.e., the set $S$ of the possible original states is all coherent states), we can choose vacuum states as the ancillas. 
When the core state is unknown, a part of the original quantum fluctuation is replaced by that of the ancillas, therefore, the added noises $n_k$ become ill-defined. 
Nonetheless, the cloner would be still optimal in some sense from the standpoint of the share of the information of $(x_\text{d},p_\text{d})$.
\end{thebibliography}
\end{document}